\documentclass[twocolumn,showpacs,superscriptaddress,prb,amsmath,amssymb,floatfix,eqsecnum]{revtex4}

\usepackage{amsmath,amssymb}
\usepackage{bm}
\usepackage{epsfig}
\usepackage{psfrag}

\newcommand{\bd}{\bm}

\begin{document}

 \title{
Quantum Heisenberg antiferromagnets in a uniform magnetic field:
Non-analytic magnetic field dependence of the magnon spectrum
}

 \author{Andreas Kreisel}  
 \affiliation{Institut f\"{u}r Theoretische Physik, Universit\"{a}t
    Frankfurt,  Max-von-Laue Strasse 1, 60438 Frankfurt, Germany}

 \author{Francesca Sauli}  
 \affiliation{Institut f\"{u}r Theoretische Physik, Universit\"{a}t
    Frankfurt,  Max-von-Laue Strasse 1, 60438 Frankfurt, Germany}

 \author{Nils Hasselmann}  
\affiliation{Institut f\"{u}r Theoretische Physik, Universit\"{a}t
    Frankfurt,  Max-von-Laue Strasse 1, 60438 Frankfurt, Germany}
 \affiliation{International Center for Condensed Matter Physics, Universidade de Bras\'ilia,\\
Caixa Postal 04667, 70910-900 Bras\'ilia, DF, Brazil}

 \author{Peter Kopietz}  
 \affiliation{Institut f\"{u}r Theoretische Physik, Universit\"{a}t
    Frankfurt,  Max-von-Laue Strasse 1, 60438 Frankfurt, Germany}

% \date{\today}
\date{April 18, 2008}

 \begin{abstract}

We re-examine the  $1/S$-correction to the
self-energy of the gapless magnon  of a $D$-dimensional
quantum Heisenberg antiferromagnet in a uniform magnetic field $h$
using a  hybrid approach between $1/S$-expansion and
non-linear sigma model,
where the  Holstein-Primakoff bosons
are expressed in terms of  Hermitian field operators
representing the uniform and the staggered components of 
the spin-operators [N. Hasselmann and P. Kopietz,
Europhys. Lett. {\bf{74}}, 1067 (2006)].
By integrating over the field associated with the uniform spin-fluctuations we
obtain the  effective action for the staggered spin-fluctuations on the lattice,
which contains fluctuations on all length scales and does not have the cutoff ambiguities
of the non-linear sigma model.
We show that in dimensions $D \leq 3$
the magnetic field dependence
of the spin-wave velocity $\tilde{c}_-( h ) $ is non-analytic in $h^2$, with
$ \tilde{c}_-( h ) - \tilde{c}_-(0 ) \propto h^2 \ln | h |$ in $D=3$, and
$ \tilde{c}_-( h ) - \tilde{c}_-(0) \propto | h |$ in $D=2$.
The frequency dependent magnon self-energy
is found to exhibit an even more singular magnetic field dependence,
implying a strong momentum dependence of the
quasi-particle residue of the gapless magnon.
We also discuss the problem of spontaneous magnon decay and show that
in $D > 1$ dimensions the damping of  magnons with momentum
$\bd{k}$ is proportional to $| \bd{k} |^{ 2 D -1 }$ if
spontaneous magnon decay is kinematically allowed.

\end{abstract}


 %\preprint{}

  %\draft

  \maketitle

\section{Introduction}

One of the most successful methods for
obtaining the low-temperature properties of ordered quantum Heisenberg magnets
is the expansion
in inverse powers of the spin quantum number
$S$. The idea is to first map the spin Hamiltonian onto
an interacting boson model
using either the Holstein-Primakoff \cite{Holstein40} or the
Dyson-Maleyev transformation \cite{Dyson56,Maleyev57}, and then
study the resulting interacting boson system
by means of the usual many-body machinery.
As the interaction vertices appearing
in the boson Hamiltonian
involve  the small parameter of $1/S$,
the perturbative
treatment of the interaction is formally justified for large $S$.
See, for example, Refs.~[\onlinecite{Oguchi60,Harris71}]
for early applications of this approach to quantum antiferromagnets (QAFM).
A disadvantage of this method is that
calculations for QAFM beyond the leading order in $1/S$
are very tedious due to a
large number of interaction vertices \cite{Harris71}.
Moreover, the vertices are even 
singular  for certain combinations of external momenta \cite{Harris71,Kopietz90,Hasselmann06}. Although the singularities cancel in physical quantities if the total spin is conserved \cite{Maleyev00},
the appearance of singularities at intermediate stages of the calculation indicates that this approach is not always the best way of calculating fluctuation
corrections to the magnon spectrum.

\begin{figure}[tb]    
   \centering
  \vspace{7mm}
    \epsfig{file=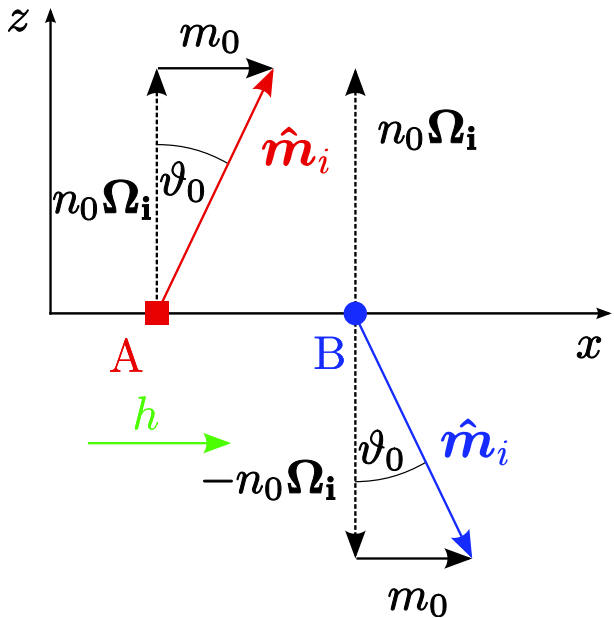,width=80mm}
  \vspace{4mm}
  \caption{%
(Color online) Spin configuration $\langle \bd{S}_i \rangle  = S \hat{\bd{m}}_i$ 
in the classical ground state
of a two-sublattice antiferromagnet subject to a uniform magnetic field
$\bd{h} = h \bd{e}_x$ in the $x$-direction.
The hypercubic lattice can be divided into two sublattices, labeled A and B,
such that the nearest neighbors of a given site all belong to the other
sublattice. The solid square denotes a site of the A-sublattice and a solid circle denotes a site of the B-sublattice.
Here  $\vartheta_0$ is the classical canting angle 
between the direction of the staggered magnetization $\bd{\Omega}_i$ and the
local spin direction $\hat{\bd{m}}_i$.
}
    \label{fig:spinconfig}
  \end{figure}

In this work we shall re-consider the leading $1/S$-correction to the
magnon self-energy of spin-$S$ quantum
Heisenberg antiferromagnets in a 
uniform magnetic field $\bd{h}$ at zero temperature in the regime where
the system  has a finite staggered magnetization.
Our starting point is the Heisenberg Hamiltonian
\begin{equation}
\hat{H} = 
 \frac{1}{2} 
\sum_{ij} J_{ij} \bd{S}_i  \cdot \bd{S}_j  -
 \sum_i \bd{h} \cdot 
 \bd{S}_i  ,
 \label{eq:hamiltonian}
 \end{equation}
where
$\bd{S}_i$ are spin operators normalized such that
$\bd{S}_i^2 = S ( S+1)$ and the magnetic field
$\bd{h}$ is measured in units of energy.  The exchange integrals
$J_{ij} $  connect nearest neighbor sites ${\bd{r}}_i$ and
$\bd{r}_j$ on 
a $D$-dimensional hypercubic lattice
with lattice spacing $a$, total volume $V = a^D N$ and $N$ sites.
As long as $|\bd{h}|$ is smaller than a certain critical value $h_c$ (see Eq. (\ref{eq:Delta0def})  below),
the spin configuration in the ground state is canted, as shown
in Fig.~\ref{fig:spinconfig}.
We choose our coordinate system such that the magnetic field
$\bd{h} = h {\bd{e}}_x$ points along the $x$-axis and the
staggered magnetization $\bd{M}_s = M_s {\bd{e}}_z$ points in $z$-direction.
The magnetic field generates 
a uniform magnetization $\bd{M} = M {\bd{e}}_x$ pointing in the same direction
as $\bd{h}$, giving via $\bd{h}$ a gap in the transverse magnon
polarized parallel to $\bd{h}$, while the magnon polarized
perpendicular to $\bd{h}$ remains gapless.

Due to the canting of the spins, the effective boson Hamiltonian
obtained from Eq.~(\ref{eq:hamiltonian})
within the Holstein-Primakoff transformation 
contains cubic interaction vertices proportional to
$S^{-1/2}$. 
Hence, to obtain the complete $1/S$-correction to physical observables,
the cubic vertices should be treated in second order perturbation theory.
The leading $1/S$-corrections to the magnon spectrum
turns out to be rather peculiar:
Zhitomirsky and Chernyshev \cite{Zhitomirsky99} 
have shown  that for  intermediate magnetic fields in a certain 
range $ h_{\ast} < |\bd{h}| < h_c$  there are
no well-defined magnons in a large part of the 
Brillouin zone due to spontaneous two-magnon decays.
Moreover,  Syromyatnikov and Maleyev \cite{Syromyatnikov01}
calculated the $1/S$-correction to the anisotropy induced gap of the magnon polarized
parallel to the magnetic field, and showed
that in dimensions $ D \leq 3 $ the correction is 
unexpectedly large. They suggested that meaningful results 
can only be obtained if  the $1/S$-expansion is re-summed to all orders, which
is of course impossible in practice.

Unfortunately, within the conventional $1/S$ expansion, the  expressions for the magnon
self-energies (see 
Refs.~[\onlinecite{Zhitomirsky99,Syromyatnikov01}]) 
are quite complicated.
For example, from the expression for the magnon self-energy given by 
Zhitomirsky and Chernyshev \cite{Zhitomirsky99} (which we reproduce in Appendix~B)  it is 
not immediately obvious
that one of the magnon branches remains gapless.
In this work we shall  therefore re-consider
this problem using our recently proposed  parameterization of the
$1/S$-expansion in terms of  Hermitian field operators \cite{Hasselmann06}.
The advantages  of such an approach
have already been pointed out  
in Ref.~[\onlinecite{Hasselmann06}], but the
practical usefulness of this method has
not been demonstrated. 
In a sense, our method is a 
hybrid approach between the $1/S$-expansion and the
non-linear sigma model (NLSM) approach \cite{Chakravarty89,Fisher89,Sachdev99}.
Recall that the NLSM is an effective
continuum theory for the staggered spin-fluctuations of a QAFM.
In contrast to the singular interaction vertices encountered in the
conventional $1/S$-expansion, the vertices describing 
interactions between transverse spin-fluctuations 
in the NLSM are finite in momentum space and all
scale as $\bd{k}^2$ for $ \bd{h} =0$.
On the other hand, 
the NLSM has to be regularized using
an ultraviolet cutoff, so that the NLSM approach  cannot be used to 
obtain the numerical value of observables which
receive contributions from wave-vectors
in the entire Brillouin zone.
Our approach combines the advantages
of the $1/S$-expansion with the those of the NLSM by parameterizing
the degrees of freedom in the $1/S$-expansion from the beginning
in terms of a lattice version of the continuum field
representing staggered spin fluctuations in the NLSM.

The rest of this work is organized as follows:
After giving a detailed description of our
hybrid approach in  Sec.~\ref{sec:HPaction}, we derive  
the effective action for staggered spin fluctuations of our lattice model  in Sec.~\ref{sec:Gausseff} and
exhibit the precise connection with the NLSM, where only the
leading orders in the derivatives are retained.
In particular, we show how the regular vertices 
of the NLSM emerge from the conventional $1/S$-expansion.
In Sec.~\ref{sec:frequency} we then use our method to derive expressions for the
frequency dependent part of the magnon self-energies which
 for small magnetic field $\bd{h}$
determines the dominant $\bd{h}$-dependence of the magnon dispersions.
In Sec.~\ref{sec:renormalization} the self-energy of the gapless magnon is evaluated;
in particular, we show that in dimensions
$ D \leq 3$ the fluctuation corrections to the spin-wave velocity and
the quasi-particle residue of the gapless magnon 
exhibit a non-analytic $\bd{h}$-dependence.
We also discuss the problem of spontaneous magnon decay in 
general dimensions.
After a brief summary of our results in Sec.~\ref{sec:conclusions},
we give  in Appendix~A  explicit expressions for the quartic interaction vertices
associated with two-magnon scattering in our hybrid approach.
Finally in Appendix~B we show numerically that in $D=2$ our result for the magnetic field dependency of the spin-wave velocity of the gapless magnon can also be extracted from the self-energy given   by Zhitomirsky and Chernyshev in Ref.~[\onlinecite{Zhitomirsky99}].

\section{Hybrid approach: 
combining the advantages of the $1/S$-expansion with those of the
NLSM}
\label{sec:HPaction}

\subsection{Holstein-Primakoff boson Hamiltonian}
\label{sec:HPboson}

For completeness, let us briefly recall the general  procedure for setting
up the $1/S$-expansion around  a given classical ground-state, characterized by
the  directions
$\hat{\bd{m}}_i  = \langle \bd{S}_i \rangle / |  \langle \bd{S}_i \rangle |$
of the local magnetic moments~\cite{Schuetz03}. 
Supplementing the unit vector  $\hat{\bd{m}}_i$ by two additional  
unit vectors ${\bd{e}}_i^{(1)}$ and ${\bd{e}}_i^{(2)}$ such that 
  ${\bd{e}}_i^{(1)}, {\bd{e}}_i^{(2)} , \hat{\bd{m}}_i $ 
form a right-handed orthogonal
triad of unit vectors, and defining the corresponding spherical basis vectors
${\bd{e}}^{p}_i $ $=$ $ {\bd{e}}^{(1)}_i + i
  p {\bd{e}}^{(2)}_i$, $p $ $=$ $ \pm$, we express the components of the spin operator
$\bd{S}_i  $ in terms of canonical boson operators $b_i$ and $b_i^{\dagger}$ 
using the 
Holstein-Primakoff transformation \cite{Holstein40},
 \begin{equation}
{\bd{S}}_i =
  S^{\parallel}_i \hat{\bd{m}}_i + {\bd{S}}^{\bot}_i
 = S^{\parallel}_i \hat{\bd{m}}_i +
  \frac{1}{2} \sum_{ p = \pm } S_i^{-p} {\bd{e}}^{p}_i 
 \; ,
 \label{eq:spinexpansion}
 \end{equation}
with
 \begin{subequations}
 \begin{eqnarray}
 S_i^{\parallel} & = & S - n_i \; \; , \; \; n_i = b^{\dagger}_i b_i
 \; ,
 \label{eq:HP1}
 \\
 S_i^{+} & = & \sqrt{2S} \sqrt{ 1 - \frac{n_i}{2S}} b_i
 \; ,
 \label{eq:HP2}
 \\
 S_i^{-} & = & \sqrt{2S} b_i^{\dagger} \sqrt{ 1 - \frac{n_i}{2S}} 
 \; .
 \label{eq:HP3}
\end{eqnarray}
 \end{subequations}
 Our spin Hamiltonian (\ref{eq:hamiltonian})
can then be written as the following bosonic many-body 
Hamiltonian~\cite{Spremo05}
 \begin{equation}
 \hat{H} = E^{\rm cl}_0 + \hat{H}_2^{\parallel} + \hat{H}_4^{\parallel}
 + \hat{H}^{\bot} + \hat{H}^{\prime}
 \; ,
 \label{eq:HH}
 \end{equation}
with the classical ground state energy
\begin{equation}
 E^{\rm cl}_0 =
 \frac{S^2}{2} \sum_{  ij}  J_{ij}  
    \hat{\bd{m}}_i \cdot \hat{\bd{m}}_j
     - S \sum_{i  }  {\bd{h}}  \cdot \hat{\bd{m}}_i 
    \label{eq:Ecl} ,
 \end{equation}
and
\begin{equation}
 \hat{H}_2^{\parallel}
  =
 - \frac{S}{2} \sum_{  ij}  J_{ij}  
    \hat{\bd{m}}_i \cdot \hat{\bd{m}}_j ( n_i + n_j )
     +  \sum_{i  }  {\bd{h}}  \cdot \hat{\bd{m}}_i n_i
    \label{eq:H2parallel}
 \; ,
 \end{equation}
\begin{eqnarray}
 \hat{H}_4^{\parallel}
  & = &
 \frac{1}{2} \sum_{  ij}  J_{ij}  
    \hat{\bd{m}}_i \cdot \hat{\bd{m}}_j  n_i  n_j 
    \label{eq:H4parallel}
 \; ,
 \\ 
\hat{H}^{\bot}  & = & \frac{1}{2} \sum_{  i j}  J_{ij}  
    {\bd{S}}^{\bot}_i \cdot {\bd{S}}^{\bot}_j
 \nonumber
 \\
 & = & \frac{1}{8} 
  \sum_{  i j}  \sum_ {p  p^{\prime}} J_{ij} 
 ( \bd{e}^p_i \cdot \bd{e}^{p^{\prime}}_j )  S_i^{-p} S_j^{- p^{\prime}}
    \,,
    \label{eq:Hbot}
 \end{eqnarray}
 \begin{eqnarray}
\hat{H}^{\prime}  & = &   - \sum_{  i}   {\bd{S}}^{\bot}_i
    \cdot \bigl(  {\bd{h}} -
    \sum_j J_{ij}  
    {{S}}_j^{ \parallel}   \hat{\bd{m}}_j   \bigr)
 \nonumber
 \\
 & = & 
- \sum_{ i j} J_{ij} 
 ( {\bd{S}}_i^{\bot} \cdot \hat{\bd{m}}_j ) n_j
 \nonumber
 \\
 &  & -
\sum_{  i}   {\bd{S}}^{\bot}_i
    \cdot \bigl(  {\bd{h}} -
    \sum_j J_{ij}  S \hat{\bd{m}}_j   \bigr)
\,. 
    \label{eq:Hrest}
 \end{eqnarray}
The part $\hat{H}^{\prime}$ of the Hamiltonian describes the
coupling between transverse and longitudinal spin fluctuations generated by the
uniform magnetic field.
Within the Holstein-Primakoff approach,
we expand the square roots in Eqs.(\ref{eq:HP2}) and (\ref{eq:HP3})
in powers of $S^{-1}$,
 \begin{subequations}
 \begin{eqnarray}
 S_i^{+} & = & 
   \sqrt{2S}\; \Biggl[ b_i - \frac{ n_i b_i   }{4S} + \ldots \Biggr] ,
 \label{eq:HP2a}
 \\
 S_i^{-} & = & \sqrt{2S}\; \Biggl[ b_i^{\dagger}  - \frac{ b_i^{\dagger} n_i    }{4S} + \ldots
 \Biggr]  .
 \label{eq:HP3a}
\end{eqnarray}
 \end{subequations}
The boson representation of
the operator $\hat{H}^{\bot}$ can then be written as an infinite series
of multiple-boson interactions involving even powers of boson operators, while
$\hat{H}^{\prime}$ becomes an infinite series of 
terms involving odd powers of boson operators, 
 \begin{eqnarray}
 \hat{H}^{\bot} & = &  \hat{H}^{\bot}_2 + \hat{H}^{\bot}_4 + O ( S^{-1} )
 \label{eq:Hbotexp}
 \; ,
 \\
\hat{H}^{\prime} & = &  \hat{H}_1 + \hat{H}_3 + O ( S^{-1/2} )
 \label{eq:Hprimeexp}
 \; ,
  \end{eqnarray}
where the subscripts indicate the number of boson operators.
Making the reasonable assumption that
the true spin configuration
in the ground state resembles the classical one
shown in Fig.~\ref{fig:spinconfig} (but with a
renormalized canting angle $\vartheta$), we have
 \begin{equation}
 \langle \bd{S}_i \rangle = | \langle \bd{S}_i \rangle |
 \hat{\bd{m}}_i \; , \; 
 \hat{\bd{m}}_i =   \zeta_i n \bd{e}_z
 + m \bd{e}_x 
 \; ,
 \end{equation}
where we have chosen
$\bd{h} = h {\bd{e}}_x$, and
the true canting angle $\vartheta$ is related to $n$ and $m$
via  $n = \cos \vartheta$ and $m = \sin \vartheta$.
Here  $\zeta_i$
assumes the value $+1$ on one sublattice (which we call the
A-sublattice) and $-1$ on the other sublattice (the B-sublattice).
A convenient choice of the other members of the local triad is
 \begin{equation}
 {\bd{e}}_i^{(1)} = \bd{e}_y \; \; , \; \;  {\bd{e}}_i^{(2)} 
= - \zeta_i n 
 \bd{e}_x + m \bd{e}_z
 \; .
 \label{eq:triad2}
 \end{equation}
The relevant scalar products  in this basis are
for nearest neighbor sites $i$ and $j$,
 \begin{subequations}
  \begin{eqnarray}
  \hat{\bd{m}}_i \cdot \hat{\bd{m}}_j & = & m^2 - n^2 = - \alpha
 \label{eq:mm}
 \; ,
 \\
 {\bd{e}}_i^{+} \cdot {\bd{e}}_j^{+} & = &  
{\bd{e}}_i^{-} \cdot {\bd{e}}_j^{-} =  
2 n^2
 \label{eq:epp}
 \; ,
 \\
 {\bd{e}}_i^{+} \cdot {\bd{e}}_j^{-} & = &  
{\bd{e}}_i^{-} \cdot {\bd{e}}_j^{+} =  
2 m^2
 \; ,
 \label{eq:epminus}
\\
 {\bd{e}}_i^{+} \cdot \hat{\bd{m}}_j  & = &  
- {\bd{e}}_i^{-} \cdot \hat{\bd{m}}_j = -2 i nm \zeta_i  = -  i \lambda \zeta_i ,
 \hspace{5mm}
 \label{eq:epmag}
 \\
 {\bd{h}}  \cdot \hat{\bd{m}}_i & = & h m ,
 \label{eq:hmscalar}
\end{eqnarray}
\end{subequations}
where we have defined
 \begin{eqnarray}
 \alpha & = & n^2 - m^2 = 1 - 2 m^2 = \cos ( 2 \vartheta ),
 \label{eq:alphadef}
 \\
 \lambda & = &  2 nm = \sin ( 2 \vartheta ).
 \label{eq:lambdadef}
 \end{eqnarray}
Then we obtain from Eq.~(\ref{eq:Ecl}),
 \begin{equation}
 E^{\rm cl}_0 = - N D J S^2 \alpha - N S h m,
 \label{eq:Ecl1} 
\end{equation}
and from Eq.~(\ref{eq:H2parallel}),
\begin{equation}
 \hat{H}_2^{\parallel}
  =   \frac{Z_h}{2} h_c \sum_i n_i ,
 \end{equation}
where 
 \begin{equation}
 Z_h  = 1 +  \frac{2 m \delta h}{h_c} ,
 \label{eq:Zhdef}
 \end{equation}
and  we have introduced the notation
 \begin{eqnarray}
 h_c & = & 4 D  J  S,
\label{eq:Delta0def}
 \\
\delta h & = & h- h_c m .
 \end{eqnarray}
In the classical limit $S \rightarrow \infty$ 
the exchange field $h_c m $ exactly cancels 
 the external field $h$, so that
in this limit $\delta h = 0$.
However, for finite $S$ the difference $\delta h = h - h_c m $ is finite. 
We shall show in Sec.~\ref{sec:Gausseff}
that $\delta h$ is actually of the order of $m h_c /S$.
The longitudinal part $\hat{H}^{\parallel}_4$ of the Hamiltonian
involving four boson operators is
 \begin{equation}
 \hat{H}_4^{\parallel} =  - \frac{\alpha}{2}  \sum_{ij} J_{ij} n_i n_j  
 \; ,
 \label{eq:H4parallel2}
 \end{equation}
and the leading two terms of the transverse part of the Hamiltonian are
 \begin{eqnarray}
 \hat{H}_2^{\bot} & = &  \frac{S}{4} \sum_{i j} J_{ij} [
 (\bd{e}^{+}_i \cdot \bd{e}^-_j ) \; b_i^{\dagger} b_j
 + (\bd{e}^{-}_i \cdot \bd{e}^+_j ) \; b_j^{\dagger} b_i
 \nonumber 
 \\
 & & \hspace{12mm} +
(\bd{e}^{+}_i \cdot \bd{e}^+_j ) \; b_i^{\dagger} b_j^{\dagger}
 + (\bd{e}^{-}_i \cdot \bd{e}^-_j ) \; b_j b_i
 ]
 \nonumber
 \\
 & = & 
\frac{S}{2} \sum_{i j} J_{ij} [
 m^2 
  ( b_i^{\dagger} b_j
 +  b_j^{\dagger} b_i ) +
n^2 (  b_i^{\dagger} b_j^{\dagger} +  b_j b_i )
 ]
 \; ,
 \nonumber
 \\
 & &
 \end{eqnarray}
\begin{eqnarray}
\hat{H}_4^{\bot} & = & - \frac{n^2}{8}\! \sum_{ij} J_{ij} \left[ n_i b_i b_j 
 + b_i n_j b_j + b^{\dagger}_i b^{\dagger}_j n_j + b^{\dagger}_i n_i b^{\dagger}_j \right]
 \nonumber
 \\
 &   &\hspace{-0.1cm} - \frac{m^2}{8} \!\sum_{ij} J_{ij} \left[ n_i b_i b^{\dagger}_j 
 + b_i b_j^{\dagger} n_j + b^{\dagger}_i n_i b_j + b^{\dagger}_i n_j b_j \right]
\; .
 \nonumber
 \\
 & &
 \label{eq:H4bot}
 \end{eqnarray}
Finally,
the part $\hat{H}^{\prime} $ of our effective boson Hamiltonian
describing the coupling between transverse and longitudinal fluctuations can
be written as
 \begin{eqnarray}
 \hat{H}^{\prime}  & = &    \lambda
\sum_{ij}  J_{ij}  \zeta_i S_i^{(2)}  n_j
% \nonumber
% \\
% & + & 
+ n \delta h \sum_i \zeta_i S^{(2)}_i
\, ,
 \nonumber
\\
 & &
 \label{eq:Hprime2}
 \end{eqnarray}
where we have set
$ S^{\pm}_i =  S_i^{(1)} \pm i S_i^{(2)}$, 
so that
 \begin{subequations}
 \begin{eqnarray}
 S_i^{(1)} & = & {\bd{e}}_i^{(1)} \cdot 
 \bd{S}_i  = \frac{1}{2} ( S_i^{+} + S_i^{-} )
 \; ,
 \label{eq:S1def}
 \\
 S_i^{(2)} & = & {\bd{e}}_i^{(2)} \cdot 
 \bd{S}_i  = \frac{1}{2i} ( S_i^{+} - S_i^{-} )
 \label{eq:S2def} .
 \end{eqnarray}
 \end{subequations}
The alternating factor $\zeta_i$ in Eq.~(\ref{eq:Hprime2})
indicates that this term
describes Umklapp scattering across the boundary of the antiferromagnetic
Brillouin zone. 
For our purpose it is sufficient to neglect all terms
in the expansion of Eq.~(\ref{eq:Hprimeexp}) involving five and more
boson operators, which amounts to retaining only $\hat{H}_1$ and
$\hat{H}_3$. With our choice of basis vectors these can be written as
\begin{equation}
 \hat{H}_1  =    n \delta h    \frac{  \sqrt{ 2 S}  }{2i}   
  \sum_i  \zeta_i 
   ( b_i - b^{\dagger}_i )
 \; ,
 \label{eq:H1def}
 \end{equation}
 \begin{equation}
 \hat{H}_3   = \lambda \frac{ \sqrt{ 2 S}   }{2i}  \sum_{ij} 
 J_{ij} \zeta_i ( b_i - b_i^{\dagger} ) n_j
 \; .
 \label{eq:H3def}
 \end{equation}
Let us emphasize that if we use the Dyson-Maleyev transformation \cite{Dyson56,Maleyev57} to bosonize the spin operators, we obtain a non-Hermitian transverse part $H_4^\bot$ which differs from Eq.~(\ref{eq:H4bot}) while $H_1$, $H_2$, $H_3$, and $H_4^\parallel$ are the same as above. Since the physical quantities calculated in this work are essentially determined by $H_3$ our results do not depend on whether we use the Holstein-Primakoff or the Dyson-Maleyev formalism.

\subsection{Linear spin-wave theory}

To obtain the magnon spectrum within
linear spin-wave theory, we neglect $\hat{H}_4^{\parallel}$ and $\hat{H}^{\prime}$, 
and approximate the transverse part $\hat{H}^{\bot}$ 
by its quadratic term in the expansion of the spin operators
in terms of the boson operators,
$\hat{H}^{\bot}  \approx  \hat{H}_2^{\bot}$.
We should now diagonalize the
quadratic boson Hamiltonian 
$\hat{H}_2 = \hat{H}_2^{\parallel} + \hat{H}_2^{\bot}$. 
We work in the
sublattice basis and Fourier transform the spin- and boson operators
on each sublattice separately: for sites $\bd{r}_i$ belonging to the
A-sublattice we define
 \begin{eqnarray}
 S_i^{(p)} & = & \sqrt{ \frac{2}{N}} \sum_{ \bd{k} } e^{ i \bd{k} \cdot
 \bd{r}_i } S_{A, \bd{k} }^{( p )}
 \label{eq:SAdef}
 \; ,
 \\
  b_i & = & \sqrt{ \frac{2}{N}} \sum_{ \bd{k} } e^{ i \bd{k} \cdot
 \bd{r}_i } A_{\bd{k}}
 \label{eq:akdef}
  \; ,
 \end{eqnarray}
 and for sites
 $\bd{r}_j$ belonging to the
B-sublattice,
 \begin{eqnarray}
 S_j^{(p)} & = & \sqrt{ \frac{2}{N}} \sum_{ \bd{k} } e^{ i \bd{k} \cdot
 \bd{r}_j } S_{B, \bd{k} }^{( p )}
 \label{eq:SBdef}
 \; ,
 \\
  b_j & = & \sqrt{ \frac{2}{N}} \sum_{ \bd{k} } e^{ i \bd{k} \cdot
 \bd{r}_j } B_{\bd{k}}
 \label{eq:bkdef}
  \; ,
 \end{eqnarray}
where the wave-vector
sums are over the reduced (antiferromagnetic) Brillouin zone.
The quadratic part $\hat{H}_2 = \hat{H}_2^{\parallel} +
 \hat{H}_2^{\bot}$ of our effective boson Hamiltonian becomes
 \begin{eqnarray}
 \hat{H}_2 & = & \tilde{J}_0 S \sum_{ \bd{k}} \bigl[
  Z_h ( A^{\dagger}_{\bd{k}} A_{\bd{k}} + B^{\dagger}_{\bd{k}} B_{\bd{k}} ) 
 \nonumber
 \\ 
& & \hspace{10mm}
+ n^2 \gamma_{\bd{k}} ( B_{ - \bd{k}} A_{\bd{k}}  +  
  A^{\dagger}_{ \bd{k}} B^{\dagger}_{- \bd{k}}   )
 \nonumber
 \\
&  & \hspace{10mm} +
 m^2 \gamma_{\bd{k}} (  B^{\dagger}_{  \bd{k}} A_{\bd{k}} +
 A^{\dagger}_{\bd{k}} B_{ \bd{k}} )
\bigr]
 \; ,
 \label{eq:H2a}
 \end{eqnarray}
where $\gamma_{\bd{k}} = \tilde{J}_{\bd{k}} / \tilde{J}_0 $ 
with
 \begin{equation}
 \tilde{J}_{ \bd{k}} = \frac{1}{N} \sum_{ i j } e^{ - i {\bd{k}} \cdot 
( {\bd{r}}_i - {\bd{r}}_j ) } J_{ij }.
 \end{equation}
Note that
 \begin{equation}
 \tilde{J}_0 S = 2 D JS = h_c /2.
 \end{equation}
To completely diagonalize $\hat{H}_2$ we first
introduce the symmetric and antisymmetric combinations
 \begin{equation}
 C_{\bd{k} \sigma } = \frac{1}{\sqrt{2}} \left[ {A}_{\bd{k}} + \sigma
 B_{\bd{k}} \right]
 \;  \; , \; \; \sigma = \pm 1
  \; ,
 \label{eq:cpmdef}
 \end{equation}
and then perform a Bogoliubov transformation,
 \begin{equation}
 \left( \begin{array}{c}
 C_{ \bd{k} \sigma } \\
 C^{\dagger}_{ - \bd{k} \sigma }  \end{array}
 \right) =
 \left( \begin{array}{cc}
 u_{ \bd{k} \sigma} & - \sigma v_{\bd{k} \sigma} \\
 -  \sigma v_{\bd{k} \sigma} & u_{ \bd{k} \sigma} \end{array} \right)
 \left( \begin{array}{c}
 \hat{\Psi}_{ \bd{k} \sigma } \\
 \hat{\Psi}^{\dagger}_{ - \bd{k} \sigma }  \end{array}
 \right)
 \; ,
 \label{eq:bogoliubov}
 \end{equation}
 where
 \begin{subequations}
 \begin{eqnarray}
 u_{ \bd{k} \sigma } & = & 
 \sqrt{ \frac{ Z_h + \sigma m^2 \gamma_{\bd{k}} + \epsilon_{\bd{k} \sigma} }{ 2 \epsilon_{\bd{k} \sigma}} }
 \; ,
 \label{eq:ukdef}
 \\
 v_{ \bd{k} \sigma } & = &
 \sqrt{ \frac{ Z_h + \sigma m^2 \gamma_{\bd{k}} - \epsilon_{\bd{k} \sigma} }{ 2 \epsilon_{\bd{k} \sigma}} }
 \; ,
 \label{eq:vkdef}
 \end{eqnarray}
 \end{subequations}
with
 \begin{eqnarray} 
 \epsilon_{ \bd{k} \sigma }    & = &  \Bigl[
 (  Z_h + \sigma m^2  \gamma_{\bd{k}} )^2 - ( n^2 \gamma_{\bd{k}} )^2 \Bigr]^{1/2}
 \nonumber
 \\
 & = & 
 \bigl[ Z_h  + \sigma  \gamma_{\bd{k}}  \bigr]^{1/2}
\bigl[ Z_h -
\sigma \alpha \gamma_{\bd{k}}  \bigr]^{1/2}
 \label{eq:epsilondef} . 
\hspace{10mm}
 \end{eqnarray}
Note that
 \begin{eqnarray}
 u^2_{ \bd{k} \sigma} +  v^2_{ \bd{k} \sigma} & = &
 \frac{ Z_h + \sigma m^2 \gamma_{\bd{k}} }{ \epsilon_{\bd{k} \sigma}},
 \\
  2 u_{ \bd{k} \sigma}  v_{ \bd{k} \sigma} & = & \frac{ n^2 \gamma_{\bd{k}}}{ 
\epsilon_{\bd{k} \sigma}}.
 \end{eqnarray}
Within linear spin-wave theory $\delta h =0$ and hence $Z_h = 1$, but
the factor $Z_h$ will deviate from unity if we take higher orders in $1/S$ into account.
Since the above transformations are canonical, our
magnon operators $\hat{\Psi}_{\bd{k} \sigma}$ satisfy the usual
bosonic commutation relations,
 \begin{equation} 
 [ \hat{\Psi}_{ \bd{k} \sigma} , \hat{\Psi}^{\dagger}_{ \bd{k}^{\prime}  \sigma^{\prime}} ] =
  \delta_{ \bd{k} , \bd{k}^{\prime} } \delta_{ \sigma , \sigma^{\prime} }
 \label{eq:Psicom}
 \; .
 \end{equation}
In terms of the new operators $\hat{\Psi}_{\bd{k} \sigma}$ the 
quadratic spin-wave Hamiltonian $\hat{H}_2$ is diagonal,
 \begin{equation}
 \hat{H}_2 =  \sum_{ \bd{k} \sigma} 
 E_{ \bd{k} \sigma} \Bigl[ \hat{\Psi}^{\dagger}_{ \bd{k} \sigma}
  \hat{\Psi}_{ \bd{k} \sigma} + \frac{1}{2}  \Bigr]
 + E_{0 \parallel}^{(1)} ,
 \label{eq:H2diagpsi}
 \end{equation}
with the magnon dispersions
 \begin{equation}
 E_{ \bd{k} \sigma } = \tilde{J}_0 S \epsilon_{ \bd{k} \sigma } .
\label{eq:Edispersiondef}
 \end{equation}
The constant
 \begin{equation}
 E_{0 \parallel}^{(1)} =- \frac{N}{2} Z_h \tilde{J}_0 S = - N D J S^2 \frac{Z_h}{S}
 \end{equation}
is the $1/S$-correction to the  ground state energy due to longitudinal spin fluctuations.
The total $1/S$-correction to the  ground state energy
is obtained by adding the zero-point energy
of the transverse spin-waves  to $E_{0 \parallel}^{(1)}$,
 \begin{eqnarray}
 E_{0}^{(1)} & = &      E_{0 \parallel}^{(1)} +  \frac{1}{2} \sum_{ \bd{k} \sigma} 
   E_{ \bd{k} \sigma } 
 \nonumber
 \\
 & = & - N D J S^2 \frac{C_1 ( h )}{S},
 \end{eqnarray}
with 
 \begin{equation}
 C_1 ( h ) = \frac{1}{N} \sum_{ \bd{k} \sigma} ( Z_h - \epsilon_{\bd{k} \sigma} ).
 \end{equation}
In the long-wavelength limit  
we obtain to linear order in $\delta h  =  h - h_c m $ and to quadratic order in $\bd{k}$, 
 \begin{subequations}
\begin{eqnarray}
 E_{ \bd{k} +}^2 & = & m h_c h   +  c_{+}^2 \bd{k}^2 ,
 \label{eq:Ekplussquared}
  \\
 E_{ \bd{k} -}^2 & = &  
n^2 m h_c \delta h  +  c_-^2  \bd{k}^2 . \hspace{10mm}
 \label{eq:Ekminussquared}
 \end{eqnarray}
 \end{subequations}
For small  $m$ the spin-wave velocities are
 \begin{subequations}
 \begin{eqnarray} 
 c_{+}^2 & = &  c_0^2 (1 - 3 m^2)
 \label{eq:cplusdef} ,
 \\
  c_-^2 & = &  c_0^2 ( n^2 +  2 m^3  \delta h  / h_c ),
 \label{eq:cminusdef}
 \end{eqnarray}
 \end{subequations}
where  $c_0$ is the
leading large-$S$ result for spin-wave velocity for $h=0$,
 \begin{equation}
 c_0 = 2 \sqrt{D} JSa.
 \end{equation}
At the level of linear spin-wave theory we may approximate the canting angle by its
classical value $\vartheta_0$, which is determined by 
the condition $\delta h =0$, or equivalently
 \begin{equation}
 m= \sin \vartheta_0 = h / h_c.
 \label{eq:cantingclassical} 
\end{equation}
This result can also be obtained
by minimizing the classical energy 
$E_0^{\rm cl}$ in Eq.~(\ref{eq:Ecl}).
The gap of  the dispersion $E_{\bd{k}+}$ is then simply given by $h$,
while the dispersion $E_{\bd{k}-}$ is gapless with spin-wave velocity 
 \begin{equation}
c_- = c_0 n = c_0 \sqrt{ 1 - \frac{ h^2}{h_c^2} }.
 \label{eq:cminusnull}
 \end{equation}

\subsection{Hermitian field operators} 
\label{subsec:Hermitian}

In the usual $1/S$-approach one now
substitutes the relations between the original Holstein-Primakoff bosons
$b_i$ and the magnon-operators $\hat{\Psi}_{\bd{k} \sigma}$ into
Eqs.~(\ref{eq:H4parallel2}, \ref{eq:H4bot}, \ref{eq:H1def}, \ref{eq:H3def}).
This yields rather lengthy expressions involving momentum dependent
vertices. However, if one is only interested in the transverse staggered
spin fluctuations, it is better perform another transformation which 
separates the staggered from the uniform spin fluctuations.
Therefore we
express the  magnon operators 
$\hat{\Psi}_{\bd{k} \sigma }$  
in terms of two Hermitian field operators
$\hat{X}_{\bd{k} \sigma}$ and $\hat{P}_{\bd{k} \sigma}$ achieving the natural normalization on a lattice as follows~\cite{Hasselmann06,Hasselmann07,Anderson52},
 \begin{equation}
 \hat{\Psi}_{\bd{k} \sigma} = p_{\sigma} \left[ \sqrt{ 
\frac{{\nu}_{\bd{k} \sigma}}{2} } 
 \hat{X}_{\bd{k} \sigma} + \frac{i}{\sqrt{ 2{\nu}_{\bd{k} \sigma}}}
 \hat{P}_{\bd{k} \sigma} \right],
 \label{eq:XPdef}
 \end{equation}
where the phase factors  $p_{+} = -i$ and $p_{-} =1$ 
are chosen for later convenience.
Here the dimensionless factors $\nu_{\bd{k} \sigma}$ are defined by
 \begin{equation}
 \nu_{\bd{k} \sigma} = \frac{E_{\bd{k} \sigma} }{  \Delta_{\bd{k} \sigma}}   ,
 \end{equation}
where 
 \begin{equation}
 \Delta_{\bd{k} \sigma} =  2 \tilde{J}_0 S z_{\bd{k} \sigma} = 
 h_c z_{\bd{k} \sigma}
,
 \end{equation}
and
 \begin{eqnarray}
  z_{\bd{k} \sigma} & = &
[ u_{ \bd{k}  \sigma } + v_{ \bd{k} \sigma } ]^2
 \epsilon_{\bd{k} \sigma}/2
 \nonumber
 \\ 
& =  &  [ Z_h + ( n^2 + \sigma m^2) \gamma_{\bd{k}} ] /2 .
 \label{eq:wdef}
 \end{eqnarray}
Note that $Z_h =1$ to leading order in $1/S$, so that to this order
 \begin{subequations}
 \begin{eqnarray}
 z_{\bd{k} +} & = &  ( 1 + \gamma_{\bd{k}} )/2,
 \\
  z_{\bd{k} -} & = &  ( 1 + \alpha \gamma_{\bd{k}} )/2,
 \end{eqnarray}
 \end{subequations}
where $ \alpha = n^2 - m^2$. In particular, 
for $\bd{k} \rightarrow 0$ we have $z_{\bd{k} + } \rightarrow 1$ and
$z_{ {\bd{k}} -} \rightarrow  (1 + \alpha)/2 =  n^2$.
One easily verifies the canonical commutation relations,
 \begin{eqnarray} 
 [  \hat{X}_{ \bd{k} \sigma } , \hat{P}_{ \bd{k}^{\prime}  \sigma^{\prime}} ]  =  
 i    \delta_{ \bd{k} , - \bd{k}^{\prime} }  \delta_{ \sigma , \sigma^{\prime}} .
 \label{eq:cancom1}
 \end{eqnarray}
The quadratic part of the spin-wave Hamiltonian
can then be written as
  \begin{eqnarray}
 \hat{H}_2 & = & \frac{ 1}{2} \sum_{ \bd{k} \sigma }
 \Delta_{\bd{k} \sigma} 
 \Bigl[  \hat{P}_{ - \bd{k} \sigma} \hat{P}_{\bd{k} \sigma} 
 + \nu_{\bd{k} \sigma}^2 
 \hat{X}_{ - \bd{k} \sigma} \hat{X}_{ \bd{k} \sigma} 
 \Bigr] + E_{0\parallel}^{(1)}.
 \nonumber
 \\
 & &
 \end{eqnarray}
In contrast to the lattice normalization of Eq.~(\ref{eq:XPdef})  in Ref.~[\onlinecite{Hasselmann07}]
we focused on the continuum limit to exhibit the relation with the NLSM. In that case
 a continuum normalization of the fields is more convenient,
 \begin{equation}
 \hat{\Psi}_{ \bd{k} \sigma}  =  p_{\sigma} 
 \sqrt{   \frac{ \chi_0}{ 2 V  E_{\bd{k} \sigma}} }
 \left[  E_{\bd{k} \sigma} \hat{\Pi}_{ \bd{k} \sigma} 
 + i \chi_0^{-1} \hat{\Phi}_{ \bd{k} \sigma}  \right],
 \label{eq:PiXhidef}
 \end{equation}
where $\chi_0 = (2 \tilde{J}_0  a^D )^{-1}$ is the large-$S$ limit of the
uniform transverse susceptibility for $h=0$.
The continuum fields fulfill the commutation relation
 \begin{eqnarray} 
 [  \hat{\Pi}_{ \bd{k} \sigma } , \hat{\Phi}_{ \bd{k}^{\prime}  \sigma^{\prime}} ]  =
 i   V \delta_{ \bd{k} , - \bd{k}^{\prime} }  \delta_{ \sigma , \sigma^{\prime}} .
 \label{eq:cancom2}
 \end{eqnarray}
The relation between lattice and continuum normalizations is
 \begin{eqnarray}
  \hat{\Pi}_{\bd{k} \sigma} & = & a^D
\sqrt{ \frac{N}{S  z_{\bd{k} \sigma}  } } \hat{X}_{\bd{k} \sigma} ,
 \label{eq:PiX} 
\\
  \hat{\Phi}_{\bd{k} \sigma} & = &
\sqrt{ NS   z_{\bd{k} \sigma} } \hat{P}_{\bd{k} \sigma} .
 \end{eqnarray}
Our spin-wave Hamiltonian (\ref{eq:H2diagpsi}) in continuum normalization
can be written as
 \begin{eqnarray}
 \hat{H}_2 & = & \frac{ 1}{2 V} \!\sum_{ \bd{k} \sigma }
 \Bigl[ \chi_0^{-1} \hat{\Phi}_{ - \bd{k} \sigma} \hat{\Phi}_{ \bd{k} \sigma} 
 + 
 \chi_0 E_{ \bd{k} \sigma}^2 \hat{\Pi}_{ - \bd{k} \sigma} \hat{\Pi}_{\bd{k} \sigma} 
 \Bigr] + E_{ 0 \parallel}^{(1)}.
 \nonumber
 \\
 & & 
 \end{eqnarray}
The field $\hat{\Pi}_{\bd{k} \sigma}$
corresponds precisely to the continuum field
representing transverse staggered spin fluctuations in the
non-linear sigma model \cite{Chakravarty89}.
However,  here we would like to calculate also
short-wavelength properties on a lattice, so that we shall work with the
lattice normalization (\ref{eq:XPdef}).

\subsection{Spin-wave interactions}

In order carry out  the $1/S$-expansion
using the  operators $X_{\bd{k} \sigma}$ and
$P_{\bd{k} \sigma}$ defined in
Eq.~(\ref{eq:XPdef}), we should first express the interaction part of the bosonized
Hamiltonian in terms of these operators.
To obtain the leading $1/S$-correction to linear spin-wave theory,
it is sufficient to approximate the effective bosonized Hamiltonian by
\begin{equation}
  \hat{H} \approx E^{\rm cl}_0 + \hat{H}_1 
 + \hat{H}_2 + \hat{H}_3 + \hat{H}_4
 \; ,
 \label{eq:Htrunc}
 \end{equation}
where $\hat{H}_4 = \hat{H}_4^{\parallel} + \hat{H}_4^{\bot}$.
Later we shall use the phase space path integral to derive the
effective action for staggered fluctuations.
All expressions in the Hamiltonian should therefore be symmetrized whenever powers of non-commutating operators are encountered~\cite{Schulman81, Negele88, Gollisch01}.
Only after symmetrization we may replace the field operators by
numbers. 
If  $\hat{A}_1 \hat{A}_2 \cdots \hat{A}_n$ is a product of operators consisting
of $\hat{X}_{ {\bd{k}} \sigma}$ or $\hat{P}_{\bd{k} \sigma}$ in arbitrary order,
the symmetrized product is
 \begin{equation}
 \bigl\{ \hat{A}_1  \hat{A}_2 \cdots \hat{A}_n \bigr\} \equiv 
 \frac{1}{ n!} \sum_{ P} 
\hat{A}_{P_1}  \hat{A}_{P_2} \cdots \hat{A}_{P_n},
 \label{eq:symdef} 
\end{equation}
where the sum is over all $n!$ permutations of $1, \ldots , n$.
We obtain from Eq.~(\ref{eq:H1def})
for the linear part  of the Hamiltonian, 
 \begin{eqnarray}
 \hat{H}_1  & = &  
    n \delta h  \sqrt{SN} \hat{P}_{0 -}    .
 \label{eq:H1cdefft}
 \end{eqnarray}
The part $\hat{H}_3$ in Eq.~(\ref{eq:H3def}) can be written as
 \begin{eqnarray}
 \hat{H}_3 & = & -  \sqrt{  \frac{N}{2} } \frac{  h_c \lambda }{ \sqrt{8S}}  \hat{P}_{0-}
 +  \sqrt{\frac{2}{N}} 
 \sum_{ \bd{k}_1 \bd{k}_2 \bd{k}_3 } \delta_{ \bd{k}_1 + \bd{k}_2 + \bd{k}_3 , 0}
\nonumber
 \\
& \times &
 \Biggr[
  \frac{1}{2 !} \Gamma^{ PXX}_{ - - -} ( \bd{k}_1 ; \bd{k}_2 , \bd{k}_3 )
 \bigl\{ \hat{P}_{ \bd{k}_1 -}  \hat{X}_{ \bd{k}_2 - } \hat{X}_{ \bd{k}_3 -} \bigr\}
\nonumber
 \\
 & & +  \frac{1}{2 !}
\Gamma^{ P XX}_{ - + +} ( \bd{k}_1 ; \bd{k}_2 , \bd{k}_3 )
 \hat{P}_{ \bd{k}_1 -} \hat{X}_{ \bd{k}_2 + } \hat{X}_{ \bd{k}_3 +}
 \nonumber
 \\
 & & +  \Gamma^{P X X }_{  + + -} ( \bd{k}_1 ; \bd{k}_2 ; \bd{k}_3 )
  \bigl\{  \hat{P}_{ \bd{k}_1 +}  \hat{X}_{ \bd{k}_2 +}  \bigr\}    \hat{X}_{ \bd{k}_3 - }
 \nonumber
 \\
 & & +   \frac{1}{2 !}   \Gamma^{PPP }_{   - + +} ( \bd{k}_1 ; \bd{k}_2 , \bd{k}_3 )
 \hat{P}_{ \bd{k}_1 -} \hat{P}_{ \bd{k}_2 + } \hat{P}_{ \bd{k}_3 +}
 \nonumber
 \\
 & & +    \frac{1}{3 !}   \Gamma^{PPP }_{   - - -} ( \bd{k}_1 , \bd{k}_2 , \bd{k}_3 )
 \hat{P}_{ \bd{k}_1 -} \hat{P}_{ \bd{k}_2 - } \hat{P}_{ \bd{k}_3 -}
\Biggl],
 \label{eq:H3primeres}
 \end{eqnarray}
where the vertices are 
 \begin{subequations}
 \begin{eqnarray}
 \Gamma^{ PXX }_{---} ( \bd{k}_1 ; \bd{k}_2 , \bd{k}_3 ) & = &
 \frac{ h_c \lambda }{\sqrt{8S}}  \gamma_{\bd{k}_1 }
 \label{eq:GammaPXX1} ,
 \\
\Gamma^{ P XX}_{ - + +} ( \bd{k}_1 ; \bd{k}_2 , \bd{k}_3 )
 & = & \frac{ h_c \lambda }{\sqrt{8S}} 
 \left[  \gamma_{\bd{k}_1 } -  \gamma_{\bd{k}_2 }  -    \gamma_{\bd{k}_3 }    \right],
 \hspace{10mm}
\label{eq:GammaPXX2}
 \\
 \Gamma^{P X X }_{  + + -} ( \bd{k}_1 ; \bd{k}_2 ; \bd{k}_3 )
 & = & \frac{ h_c \lambda }{\sqrt{8S}} 
  \gamma_{\bd{k}_2 },
 \label{eq:GammaPXX3}
 \\
 \Gamma^{PPP }_{   - + +} ( \bd{k}_1 ; \bd{k}_2 , \bd{k}_3 )
 & = & \frac{ h_c \lambda }{\sqrt{8S}}  \gamma_{\bd{k}_1 }
 \label{eq:GammaPPP1} ,
 \\
 \Gamma^{PPP }_{  - - -} ( \bd{k}_1 , \bd{k}_2 , \bd{k}_3 )
 & = &
 \frac{ h_c \lambda }{\sqrt{8S}}
   \left[   \gamma_{\bd{k}_1}  +   \gamma_{\bd{k}_2}    +   \gamma_{\bd{k}_3}
 \right].   
 \label{eq:GammaPPP2}
 \end{eqnarray} 
\end{subequations}
Explicitly, the symmetrized products in Eq. (\ref{eq:H3primeres})  are
 \begin{eqnarray}
 \bigl\{ \hat{P}_1 \hat{X}_2 \bigr\} & = & \frac{1}{2} \left[ \hat{P}_1 , \hat{X}_2 
 \right]_{+},
 \\
 \bigl\{ \hat{P}_1 \hat{X}_2 \hat{X}_3 \bigr\} & = &
\frac{1}{3} ( \hat{P}_1 \hat{X}_2 \hat{X}_3 +  \hat{X}_2 \hat{X}_3 \hat{P}_1 )
 \nonumber
 \\
 & + & \frac{1}{6} ( \hat{X}_2 \hat{P}_1 \hat{X}_3 +  \hat{X}_3 \hat{P}_1 \hat{X}_2 )
 \nonumber
 \\
 & = & \frac{1}{2} \left[ \hat{P}_1 , \hat{X}_2 \hat{X}_3
 \right]_{+},
 \end{eqnarray}
where $ [ \hat{A}_1 , \hat{A}_2 ]_+ =  \hat{A}_1  \hat{A}_2 + \hat{A}_2  \hat{A}_1$
is the anti-commutator and we have abbreviated 
$\hat{P}_{\bd{k}_1}$ by $\hat{P}_1$ and analogously for the other labels.
 
Finally, consider the part $\hat{H}_4 = \hat{H}_4^{\parallel} + \hat{H}_4^{\bot}$
of the Hamiltonian involving four boson operators, which according
to Eqs.~(\ref{eq:H4parallel2}) and (\ref{eq:H4bot}) is given by
\begin{eqnarray}
\hat{H}_4 & = & - \frac{n^2}{2 } \sum_{ij} J_{ij} \Biggl\{ n_i n_j 
+ \frac{1}{4} \bigl[ n_i b_i b_j 
 + b_i n_j b_j 
 \nonumber
 \\
 &  & 
\hspace{10mm} + b^{\dagger}_i b^{\dagger}_j n_j + b^{\dagger}_i n_i b^{\dagger}_j \bigr]
 \Biggr\}
 \nonumber
 \\
 &   & + \frac{m^2}{2} \sum_{ij} J_{ij} 
 \Biggl\{ n_i n_j - \frac{1}{4} \bigl[ n_i b_i b^{\dagger}_j 
 + b_i b_j^{\dagger} n_j 
 \nonumber
 \\
& & \hspace{10mm} + b^{\dagger}_i n_i b_j + b^{\dagger}_i n_j b_j \bigr]
 \Biggr\} .
 \label{eq:H4}
 \end{eqnarray}
Expressing $\hat{H}_4$ in terms of the
operators $\hat{P}_{\bd{k} \sigma}$ and $\hat{X}_{ \bd{k} \sigma}$
defined in Eq.~(\ref{eq:XPdef})
and symmetrizing all expressions containing non-commuting
operators we obtain
 \begin{equation}
 \hat{H}_4 =  E_{0 \parallel}^{(2)} + \delta \hat{H}_2^{\prime} + \hat{H}_4^{\prime},
 \label{eq:H4sum}
 \end{equation}
where 
  \begin{equation}
   E_{0 \parallel}^{(2)} = 
 -    \frac{N D J S^2 \alpha}{(2 S)^2}
 \label{eq:Eclcor}
 \end{equation}
is a $1/S^2$-correction to the classical ground state energy, and
 \begin{equation}
 \delta \hat{H}_2^{\prime} = 
 \frac{   1    }{2} 
 \sum_{\bd{k} \sigma} \Bigl[  \Gamma^{P }_{\sigma } 
( \bd{k} )   \hat{P}_{ - \bd{k}  \sigma} \hat{P}_{\bd{k} \sigma }
+  \Gamma^{X  }_{\sigma } 
( \bd{k} )   \hat{X}_{ - \bd{k}  \sigma} \hat{X}_{\bd{k} \sigma }
\Bigr],
 \label{eq:H2cor}
 \end{equation}
is a $1/S$-correction to $\hat{H}_2$. The vertices are
\begin{subequations}
 \begin{eqnarray}
 \Gamma^{P }_{\sigma} ( \bd{k} ) & = & \frac{h_c}{4S}
   \alpha (1  + \sigma \gamma_{\bd{k}}  ),
 \label{eq:GammaPhiPhi1}
 \\
 \Gamma^{X  }_{ \sigma} ( \bd{k} ) & = & \frac{h_c}{4S}
  (\alpha   - \sigma \gamma_{\bd{k}} ).
 \label{eq:GammaPiPi2}
\end{eqnarray}
\end{subequations}
Finally, the properly symmetrized quartic part $\hat{H}_4^{\prime}$ of our spin-wave 
Hamiltonian is given in  Appendix~A.
For our purpose it is only important that the corresponding interaction vertices are
non-singular functions of the external momenta and
are analytic functions of  $h^2$.

\section{Effective action for the staggered spin fluctuations}
\label{sec:Gausseff}
In Ref.~[\onlinecite{Hasselmann06}]
the precise relation between the 
magnon quasi-particle operators 
of the $1/S$-expansion and the
continuum fields $\Pi_{\bd{k}\sigma}$
representing transverse fluctuations of the staggered magnetization
has been established. 
In this section we shall use this relation to derive the effective action
for the staggered spin fluctuations for 
the Hamiltonian (\ref{eq:hamiltonian}) retaining 
sub-leading $1/S$-corrections and short wave length
fluctuations in the entire Brillouin zone.

\begin{figure}[tb]    
   \centering
  \vspace{7mm}
      \epsfig{file=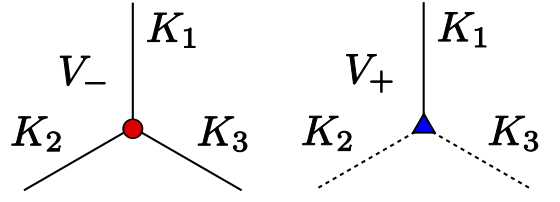,width=70mm}
  \vspace{4mm}
  \caption{%
(Color online) Graphical representation of the interaction vertices $V_{+} ( K_1, K_2 , K_3 )$
and $V_{-} ( K_1, K_2 , K_3 )$ defined in Eqs.~(\ref{eq:Vminusdef}) and
(\ref{eq:Vplusdef}). Solid lines represent  
the gapless field $X_{-}$, while dashed lines correspond to the
gapped field $X_{+}$.
The shape of the symbols reflects the
symmetry of the vertices with respect to the permutation of the labels.}
    \label{fig:vertices}
  \end{figure}

For weak magnetic fields, the operators $\hat{P}_{\sigma}$ 
correspond to transverse fluctuations of the total spin, while
  $\hat{X}_{\sigma}$ describe staggered (antiferromagnetic) spin fluctuations.
To calculate the self-energy of antiferromagnetic magnons,
we can therefore eliminate the degrees of freedom associated with the generalized
momenta $\hat{P}_{\sigma}$. This is most conveniently done using path
integration.
The appropriate path integral in our case is the imaginary time phase space path 
integral \cite{Schulman81,Negele88}.  Recall that for a one-dimensional quantum mechanical
system with 
position operator $\hat{X}$, momentum operator $\hat{P}$, and
Hamiltonian $\hat{H} ( \hat{P} , \hat{X} )$ the partition function
can be written as
 \begin{equation} 
 {\cal{Z}} = \int {\cal{D}} [P ,  X ]
 \exp \left\{ \int_0^{\beta} d \tau \left[ i P \frac{\partial X }{ \partial \tau} -  H_s ( P , X )   \right]
 \right\}
 \; ,
 \label{eq:phaseint}
 \end{equation}
where $H_s ( P , X )$ is obtained from the
Hamiltonian $\hat{H} ( \hat{P} , \hat{X} )$ by first symmetrizing
$\hat{H} ( \hat{P} , \hat{X} )$ with respect to the ordering of the operators
$\hat{X}$ and $\hat{P}$, and then replacing the
operators by their eigenvalues. In principle,
ambiguities associated with the
operator ordering in the phase space
path integral can always be resolved by
going back to the discretized definition of the path integral \cite{Schulman81,Negele88}.
However, recently Gollisch and Wetterich~\cite{Gollisch01, Wetterich07} showed 
that in the continuum notation the  symmetrization prescription 
leads to the same result as the more fundamental 
discretized definition of the phase space path integral.
The Euclidean action corresponding to our spin-wave Hamiltonian
is of the form
 \begin{equation}
 S [ P_{\sigma} , X_{\sigma} ]
 = \sum_{ l=0}^{\infty}
 S_l [ P_{\sigma} , X_{\sigma} ] 
 \; ,
 \label{eq:series}
 \end{equation}
where $S_l [ P_{\sigma} , X_{\sigma} ] $ contains
$l$ powers of the fields.
To obtain the effective action 
$S_{\rm eff} [ X_\sigma ]$ for
the staggered fluctuations, we 
integrate over the generalized momenta,
 \begin{equation}
 e^{ - S_{\rm eff} [ X_{\sigma} ] }
 =   \int {\cal{D}} [ P_{\sigma} ] 
 e^{ - S [ P_{\sigma} , X_{\sigma} ]  } .
 \label{eq:SeffXdef}
 \end{equation}
Within the Gaussian approximation (corresponding  to linear spin-wave theory)
we truncate the expansion (\ref{eq:series})
at the term $l=2$. The relevant contributions to 
 $S [ P_{\sigma} , X_{\sigma} ]$ can be written as
 \begin{eqnarray}
 S_0 & = &  \beta  [  E^{\rm cl}_0 + E_{0 \parallel}^{(1)} ] ,
 \label{eq:S0def}
 \\ 
S_1 [ P_- ] & = & \beta n \delta h  \sqrt{SN} P_{0-},
 \label{eq:S1action}
 \end{eqnarray}
and
 \begin{eqnarray}
  S_2 [ P_{\sigma} , X_{\sigma} ] & = & 
 \frac{\beta}{2} 
 \sum_{K , \sigma}  \Bigl[ \Delta_{\bd{k} \sigma} \left(
P_{ - K \sigma} P_{K \sigma} 
+ \nu_{\bd{k} \sigma}^2     X_{ - K \sigma} X_{K \sigma}
\right)
 \nonumber 
 \\
 &  &  \hspace{5mm}  -  \omega  ( P_{ - K \sigma} X_{K \sigma} 
 - X_{ - K \sigma} P_{K \sigma} ) 
  \Bigr] ,
 \label{eq:S2pp}
 \end{eqnarray}
where the last term in Eq.~(\ref{eq:S2pp}) corresponds to the measure term 
$i P \partial X / \partial \tau$
in the
phase space functional integral (\ref{eq:phaseint}).
The fields $P_{ K \sigma}$ and 
$X_{ K \sigma}$ are defined by
replacing the  operators
$\hat{P}_{ \bd{k}  \sigma } $ and $ \hat{X}_{ \bd{k} \sigma }$ by quantum fields
$P_{ \bd{k}  \sigma } ( \tau )$ and $ X_{ \bd{k} \sigma } ( \tau )$
depending on imaginary time $\tau$
and expanding  the fields in frequency space,
 \begin{subequations}
 \begin{eqnarray}
 P_{ \bd{k}  \sigma } ( \tau ) & = &  \sum_{\omega} e^{ - i \omega \tau } 
P_{ K \sigma }  ,
 \label{eq:PiMatsubara}
 \\
X_{ \bd{k}  \sigma } ( \tau ) & = &  \sum_{\omega} e^{ - i \omega \tau } 
X_{ K \sigma }.
 \label{eq:PhiMatsubara}
 \end{eqnarray}
\end{subequations}
We combine
momenta $\bd{k}$ and bosonic Matsubara frequencies $i \omega$ 
to form  a composite label $K = ( \bd{k} , i \omega)$.
In general the canting angle can be determined from the condition
that the functional average of the field $P_{K=0,-}$ vanishes,
 \begin{equation}
 \langle P_{ 0-} \rangle =0.
 \label{eq:cantingren}
 \end{equation}
Eq.~(\ref{eq:cantingren}) defines the correction $\delta h = h - h_c m = h - h_c \sin \vartheta$ and hence the sine of the renormalized canting angle $\sin \vartheta = m = (h - \delta h)/h_c$. 
Within the Gaussian approximation this  implies
$\delta h =0$, leading to the classical result (\ref{eq:cantingclassical}).
Hence $S_1 [ P_- ] =0$ within this approximation and the
effective action for the fields $X_{\sigma}$ is  given by the Gaussian integral
 \begin{equation}
 e^{ - S_{{\rm eff}} [ X_{\sigma} ] }
 \approx  e^{ -S_0}
 \int {\cal{D}} [ P_{\sigma}   ] 
e^{ - S_2 [ P_{\sigma} , X_{\sigma} ]} .
 \label{eq:Zgauss}
 \end{equation}
Carrying out the integration, we obtain
in Gaussian approximation $S_{{\rm eff}} [ X_{\sigma} ] = S_0
+ S_{{\rm eff}}^{(0)} [ X_{\sigma} ]$,
where
 \begin{eqnarray} 
S_{{\rm eff}}^{(0)} [ X_{\sigma} ]  & = &
 \frac{\beta}{2} \sum_{K  \sigma}
 \frac{ E_{\bd{k} \sigma}^2 + \omega^2 }{   \Delta_{\bd{k} \sigma } }
 X_{ -K \sigma} X_{ K \sigma} .
 \label{eq:P2Pires}
 \end{eqnarray}
At long wavelengths this action has the same form as the
corresponding Gaussian part of the action of the NLSM.
However, in contrast to the NLSM, our action is defined on the lattice so that
fluctuations on all wavelengths are included.
The Gaussian propagator of the $X_{\sigma}$-field is thus
 \begin{equation}
 \langle X_{ K \sigma} X_{ K^{\prime} \sigma^{\prime} } \rangle_0 =
 \delta_{ K , - K^{\prime}  } \delta_{\sigma \sigma^{\prime}}
( \beta  \Delta_{\bd{k} \sigma } )^{-1}
 \frac{  \Delta_{\bd{k} \sigma }^2 }{  E_{\bd{k} \sigma}^2 + \omega^2  }.
 \label{eq:Xprop}
 \end{equation}
The other propagators are within Gaussian approximation
 \begin{eqnarray}
 \langle P_{ K \sigma} P_{ K^{\prime} \sigma^{\prime} } \rangle_0 & = &
 \delta_{ K , - K^{\prime}  } \delta_{\sigma \sigma^{\prime}}
( \beta  \Delta_{\bd{k} \sigma } )^{-1}
\frac{   E_{\bd{k} \sigma }^2 }{   
 E_{\bd{k} \sigma}^2 + \omega^2  }, \hspace{7mm}
 \label{eq:Pprop}
 \\
 \langle X_{ K \sigma} P_{ K^{\prime} \sigma^{\prime} } \rangle_0 & = &
 \delta_{ K , - K^{\prime}  } \delta_{\sigma \sigma^{\prime}}
( \beta  \Delta_{\bd{k} \sigma } )^{-1}
 \frac{  \Delta_{\bd{k} \sigma }   \omega  }{  
 E_{\bd{k} \sigma}^2 + \omega^2  }.
 \label{eq:PXprop}
 \end{eqnarray}
Here the symbol $ \langle \ldots \rangle_0$ denotes functional averaging
with  the Gaussian action $S_2 [P_{\sigma} , X_{\sigma} ]$. 
Note that the formal sum $\sum_{\omega} \langle X_{ K \sigma} P_{ -K \sigma} 
 \rangle_0 $ represents the expectation value of the
symmetric operator $ \langle \{ \hat{X}_{\bd{k} \sigma} \hat{P}_{\bd{k} \sigma} \} \rangle_0 =0$,
so that we should regularize formally divergent Matsubara sums
using a symmetric convergence factor $\cos ( \omega 0^+ )$, 
 \begin{equation}
 \langle \{ \hat{X}_{\bd{k} \sigma} \hat{P}_{\bd{k} \sigma} \} \rangle_0
 = \frac{1}{\beta} \sum_{\omega} 
\frac{   \omega  \cos ( \omega 0^+ )   }{  
 E_{\bd{k} \sigma}^2 + \omega^2  }     =0.
 \end{equation}
The higher  $1/S$-corrections
to $S_{\rm eff} [ X_{\sigma} ]$,
can now be obtained by including the
spin-wave interactions perturbatively.
Therefore we rewrite Eq.~(\ref{eq:SeffXdef}) as
 \begin{equation}
 S_{\rm eff} [ X_{\sigma} ] = S_0 + S_{{\rm eff}}^{(0)} [ X_{\sigma} ] + 
S_{\rm eff}^{\rm int} [ X_{\sigma} ],
\end{equation}
where the interaction part $S_{\rm eff}^{\rm int} [ X_{\sigma} ]$ is defined via the
following functional average,
 \begin{eqnarray}
 S_{\rm eff}^{\rm int} [ X_{\sigma} ] & = & - \ln 
\left\langle e^{ - S_{\rm int} [ P_{\sigma} , X_{\sigma} ] } \right\rangle_{P}
\nonumber
 \\
 &  & \hspace{-20mm} \equiv - \ln \left[
 \frac{ \int {\cal{D}} [ P ] e^{ - S_2 [ P_{\sigma} , X_{\sigma} ] } e^{ -  
 S_{\rm int} [ P_{\sigma} , X_{\sigma} ] } }{
\int {\cal{D}} [ P ] e^{ - S_2 [ P_{\sigma} , X_{\sigma} ] } }
 \right],
 \end{eqnarray} 
where
 \begin{equation}
 S_{\rm int} [ P_{\sigma} , X_{\sigma} ]  = S_1 [ P_- ] + 
\sum_{l=3}^{\infty} S_{l} [ P_{\sigma} , X_{\sigma} ].
 \end{equation}
The leading correction of relative order $1/S$ arises from the
first order correction due  to $S_4 [ P_{\sigma} , X_{\sigma} ] $ corresponding to
$\hat{H}_4$ 
defined in  Eqs.~(\ref{eq:H4sum}, \ref{eq:Eclcor}, \ref{eq:H2cor}, \ref{eq:h4vertex}),
and the
second order corrections due to the sum of  
$S_1 [ P_-]$ and $S_3 [ P_{\sigma} , X_{\sigma} ]$, corresponding to
$\hat{H}^{\prime} \approx \hat{H}_1 + \hat{H}_3$ in
Eqs.~(\ref{eq:H1cdefft}) and (\ref{eq:H3primeres}).
Note that to order $1/S$ the difference $\delta h = h - h_c m$
and hence $S_1 [ P_- ]$ are finite, so that
the condition (\ref{eq:cantingren}) for the renormalized
canting angle reduces to
 \begin{equation}
 \Bigl\langle P_{0-}  \bigl( S_1 [ P_{0-} ]  +
 S_3 [ P_{\sigma} , X_{\sigma} ] \bigr) \Bigr\rangle_0 =0.
 \label{eq:PPcant} 
\end{equation}
Performing the Gaussian averages we obtain 
to first order in $1/S$,
 \begin{equation}
 \delta h  =  m  [ 1- C_2 ( h )] \frac{h_c}{ 2S}
 \label{eq:lambdares}  ,
 \end{equation}
with the numerical constant
 \begin{eqnarray}
 C_2 ( h )  & = &  \frac{1}{N} \sum_{ \bd{k} \sigma}
 \left[ u_{\bd{k} \sigma}^2 + v_{ \bd{k} \sigma}^2 - \sigma \gamma_{\bd{k}}
 ( u_{\bd{k} \sigma} + \sigma v_{\bd{k} \sigma } )^2 \right]
 \nonumber
 \\
& = & \frac{1}{N}\sum_{ \bd{k} \sigma} 
  \frac{ 1 - \gamma_{\bd{k}}^2 - \sigma n^2 \gamma_{\bd{k}} }{
 \epsilon_{\bd{k} \sigma} } 
 \label{eq:I1def} .
  \end{eqnarray}
Our condition (\ref{eq:lambdares}) leads to the same $1/S$-corrections for the canting angle 
as in Ref.~[\onlinecite{Zhitomirsky98}] and thus yields the same result for the uniform magnetization.
Note that $S_1 [ P_- ]$ is of order $S^{-1/2}$ and should be taken into account 
on the same footing with $S_3 [ P_{\sigma} , X_{\sigma} ]$
in  second order  perturbation theory to collect all corrections of relative order
$1/S$. Using Eq.~(\ref{eq:lambdares}) 
we obtain for the total contribution of order $S^{-1/2}$
to the action $S^{\prime} [ P_{\sigma} , X_{\sigma} ]$ 
corresponding to $\hat{H}^{\prime}$ in Eq.~(\ref{eq:Hprimeexp}),
 \begin{eqnarray}
 S^{\prime} [ P_{\sigma} , X_{\sigma} ] &   \approx & S_1 [ P_{-} ] + 
S_3  [ P_{\sigma} , X_{\sigma} ] 
 \nonumber
 \\
 &  & \hspace{-20mm}  = - \beta  \sqrt{  \frac{N}{2} } \frac{  h_c \lambda }{ \sqrt{8S}}  
 C_2 ( h )  {P}_{0-}
\nonumber
 \\
 &  & \hspace{-20mm} + \beta \sqrt{\frac{2}{N}} 
 \sum_{ K_1 K_2 K_3 } \delta_{ K_1 + K_2 + K_3 , 0}
\nonumber
 \\
&  & \hspace{-18mm} \times
 \Biggr[
  \frac{1}{2 !} \Gamma^{ PXX}_{ - - -} ( {\bd{k}}_1 ; {\bd{k}}_2 , {\bd{k}}_3 )
  {P}_{ K_1 -} {X}_{ {K}_2 - } {X}_{ {K}_3 -} 
\nonumber
 \\
 & & \hspace{-15mm} +  \frac{1}{2 !}
\Gamma^{ P XX}_{ - + +} ( \bd{k}_1 ; \bd{k}_2 , \bd{k}_3 )
 {P}_{ {K}_1 -} {X}_{ {K}_2 + } {X}_{ {K}_3 +}
 \nonumber
 \\
 & &  \hspace{-15mm}   +  \Gamma^{P X X }_{  + + -} ( \bd{k}_1 ; \bd{k}_2 ; \bd{k}_3 )
  {P}_{ {K}_1 +}  {X}_{ {K}_2 +}    {X}_{ {K}_3 - }
 \nonumber
 \\
 & & \hspace{-15mm}   
+   \frac{1}{2 !}   \Gamma^{PPP }_{   - + +} ( \bd{k}_1 ; \bd{k}_2 , \bd{k}_3 )
 {P}_{ {K}_1 -} {P}_{ {K}_2 + } {P}_{ {K}_3 +}
 \nonumber
 \\
 & & \hspace{-15mm}    +  
  \frac{1}{3 !}   \Gamma^{PPP }_{   - - -} ( \bd{k}_1 , \bd{k}_2 , \bd{k}_3 )
 {P}_{ {K}_1 -} {P}_{ {K}_2 - } {P}_{ {K}_3 -}
\Biggl].
 \label{eq:H3primeres2}
 \end{eqnarray}
The leading correction to the Gaussian approximation
for the effective action $S_{\rm eff} [ X_{\sigma} ]$
is of  order $1/ \sqrt{S}$,
 \begin{eqnarray}
 S_{ {\rm eff}}^{ (1/2)} [ X_{\sigma} ]  & = &  \langle S^{\prime} 
[ P_{\sigma} , X_{\sigma} ]  \rangle_P,
 \label{eq:Shalf}
 \end{eqnarray}
where the subscript indicates the power of $1/S$.
The   $1/S$-correction is
 \begin{eqnarray}
 S_{ {\rm eff}}^{ (1)  } [ X_{\sigma} ]  & = &  
\langle S_4 [ P_{\sigma} , X_{\sigma} ]  \rangle_P  
 \nonumber
 \\
 &  & \hspace{-30mm} - \frac{1}{2}  
 \Big\langle 
\bigl(  S^{\prime} [ P_{\sigma} , X_{\sigma} ]     - \langle 
  S^{\prime} [ P_{\sigma} , X_{\sigma} ] \rangle_P       \bigr)^2 \Big\rangle_P .
 \end{eqnarray}
To calculate the Gaussian average in Eq.~(\ref{eq:Shalf}) we use the fact
that averaging  the field $P_{K \sigma}$ for fixed $X$ yields
 \begin{equation}
 \langle P_{ K \sigma} \rangle_P = \frac{\omega}{\Delta_{\bd{k} \sigma} } X_{ K \sigma}.
 \end{equation}
After proper symmetrization of the vertices we obtain
 \begin{eqnarray}
 S_{ {\rm eff}}^{ (1/2)} [ X_{\sigma} ] & = &  
  \beta \sqrt{ \frac{2}{N}}
 \sum_{ K_1  K_2 K_3} \delta_{ K_1 + K_2 + K_3,0} 
 \nonumber
 \\
& & \hspace{-20mm} \times
\Bigl[  \frac{1}{3!}   \Gamma^{(3)}_{ ---} ( K_1 , K_2 , K_3 )        
X_{ K_1 -} X_{ K_2 - }  X_{ K_3 - }
 \nonumber
 \\
& & \hspace{-18mm} + \frac{1}{2!} \Gamma^{(3)}_{ -++} ( K_1 ; K_2 , K_3 )
X_{ K_1 -} X_{ K_2 + }  X_{ K_3 + }
\Bigr], \hspace{10mm}
 \label{eq:Seffhalf}
 \end{eqnarray}
with
 \begin{eqnarray}
  \Gamma^{(3)}_{ ---} ( K_1 , K_2 , K_3 ) & = & \frac{ h_c \lambda}{\sqrt{8S}}
 \Bigl[     \frac{   \gamma_{ \bd{k}_1} \omega_1   }{ \Delta_{ \bd{k}_1-}}  
+ \frac{  \gamma_{ \bd{k}_2}  \omega_2 }{ \Delta_{ \bd{k}_2-}}
+ \frac{   \gamma_{ \bd{k}_3}  \omega_3 }{ \Delta_{ \bd{k}_3-}}
 \nonumber
 \\
 &  & \hspace{-10mm} + \frac{ 
( \gamma_{ \bd{k}_1} +     \gamma_{ \bd{k}_2}  + \gamma_{ \bd{k}_2} ) 
\omega_1   \omega_2 \omega_3     }{
 \Delta_{ \bd{k}_1-}    \Delta_{ \bd{k}_2-}  \Delta_{ \bd{k}_3-}}
\Bigr],
 \label{eq:Gamma3a}
 \end{eqnarray}    
\begin{eqnarray}
  \Gamma^{(3)}_{ -++} ( K_1 ; K_2 , K_3 ) & = & \frac{ h_c \lambda}{\sqrt{8S}}
 \Bigl[    (   \gamma_{ \bd{k}_1}
- \gamma_{ \bd{k}_2} - \gamma_{ \bd{k}_3} )  \frac{ \omega_1   }{ \Delta_{ \bd{k}_1-}} 
\nonumber
 \\
 &  & \hspace{-25mm}
+ \frac{   \gamma_{ \bd{k}_2} \omega_3 }{ \Delta_{ \bd{k}_3+}}
+ \frac{  \gamma_{ \bd{k}_3} \omega_2 }{ \Delta_{ \bd{k}_2+}}
  + \frac{ \gamma_{ \bd{k}_1}   \omega_1   \omega_2 \omega_3     }{
 \Delta_{ \bd{k}_1-}    \Delta_{ \bd{k}_2+}  \Delta_{ \bd{k}_3+}}
\Bigr].
 \label{eq:Gamma3b}
 \end{eqnarray}    
Actually, the terms cubic in the frequencies which are due to the  cubic
terms in the $P_{K \sigma}$ in Eq.~(\ref{eq:H3primeres2})
 can be omitted, because the contribution of these terms to the self-energy
of the $X$-fields is frequency-independent to order $1/S$. 
Since we are only interested in the frequency dependent part of the
self-energy, 
we may thus replace
 \begin{eqnarray}
 \Gamma^{(3)}_{ ---} ( K_1 , K_2 , K_3 )   & \rightarrow &  V_{-} ( K_1 , K_2, K_3 )
 \nonumber
 \\ 
 &  & \hspace{-35mm} \equiv 
 \frac{ h_c \lambda}{\sqrt{8S}}
 \Bigl[      
  \frac{   \gamma_{ \bd{k}_1} \omega_1   }{ \Delta_{ \bd{k}_1-}}  
+ \frac{  \gamma_{ \bd{k}_2}  \omega_2 }{ \Delta_{ \bd{k}_2-}}
+ \frac{   \gamma_{ \bd{k}_3}  \omega_3 }{ \Delta_{ \bd{k}_3-}}
\Bigr], \hspace{10mm}
 \label{eq:Vminusdef}
 \\
\Gamma^{(3)}_{ -++} ( K_1 ; K_2 , K_3 )   & \rightarrow &  V_{+} ( K_1 , K_2, K_3 )
 \nonumber
 \\ 
 &  & \hspace{-35mm} \equiv 
\frac{ h_c \lambda}{\sqrt{8S}}
 \Bigl[      ( \gamma_{ \bd{k}_1}    - \gamma_{ \bd{k}_2} - \gamma_{ \bd{k}_3} )
\frac{ \omega_1   }{ \Delta_{ \bd{k}_1-}} 
+ \frac{   \gamma_{ \bd{k}_2} \omega_3 }{ \Delta_{ \bd{k}_3+}}
+ \frac{  \gamma_{ \bd{k}_3} \omega_2 }{ \Delta_{ \bd{k}_2+}}
\Bigr].
 \nonumber
 \\
 & & 
 \label{eq:Vplusdef}
 \end{eqnarray}
Graphical representations of the interaction vertices $V_{\sigma} ( K_1, K_2 , K_3 )$
are shown in Fig.~\ref{fig:vertices}.

At this point we can make contact with the NLSM, which is an effective low-energy
theory for staggered spin fluctuations.
In the presence of a uniform magnetic field the Euclidean action of the
NLSM  is \cite{Sachdev99,Fisher89},
 \begin{eqnarray}
S_{\rm{NLSM} }
[ \bd{\Omega} ]  & = & \frac{ \rho_s}{2} \int_0^{\beta} d \tau
 \int d^{D} r \Bigl[ \sum_{ \mu =1}^{D} ( \partial_{\mu} \bd{\Omega})^2 
\nonumber
 \\
 & & \hspace{10mm}
 + c^{-2} ( \partial_{\tau}  \bd{\Omega}  - i \bd{h} \times  
 \bd{\Omega})^2 \Bigr] , \hspace{10mm}
 \label{eq:sigmah} 
 \end{eqnarray}
where the unit vector  $\bd{\Omega} ( \tau , \bd{r} )$ 
represents the slowly  fluctuating staggered magnetization,
$\rho_s$  and $c$ are the  spin stiffness and the spin-wave velocity
at temperature $T=0$, and
$\partial_{\mu} = \partial / \partial r_{\mu}$ is the spatial
derivative in direction $\mu = 1, \ldots , D$.
The model (\ref{eq:sigmah}) can be obtained from the
corresponding NLSM for $\bd{h} =0$ 
by substituting
 $ \partial_{\tau}  \to \partial_{\tau}  - i \bd{h} \times$.
Although this procedure 
does not explicitly take into account the magnetic field dependence of 
the spin-wave velocity and
the spin stiffness, one usually argues that $c$ and $\rho_s$
in Eq.~(\ref{eq:sigmah}) are effective parameters, implicitly including the
effect of the magnetic field. However, this procedure is based on the
assumption that in the presence of a magnetic field the
magnon dispersions can be characterized by a single spin-wave velocity
$c ( h )$. From Eqs.~(\ref{eq:cplusdef}) and (\ref{eq:cminusdef}) 
it is clear that this assumption is not justified, because the dispersion of
spin-wave mode polarized parallel to the magnetic field
involves a different spin-wave velocity  than the mode
polarized perpendicular to the magnetic field \cite{Hasselmann07}.
Apparently, there are no published calculations of the
$1/S$-corrections to the magnetic field dependence of
the spin-wave velocity.
In the following section we shall show
that in dimensions $D \leq  3$ the magnetic field dependence of
the spin-wave velocity $c_{-} ( h )$
of the gapless magnon mode is non-analytic in $h^2$.

To make contact with our spin-wave approach, let us consider
the interaction vertex due to the magnetic field in the NLSM.
Therefore we rewrite Eq.~(\ref{eq:sigmah}) as
 \begin{eqnarray}
S_{\rm{NLSM} }
[ \bd{\Omega} ] 
 & =  & 
 \frac{ \rho_s}{2} \int_0^{\beta}\! d \tau
 \int\! d^{D} r \Bigl[ \sum_{ \mu =1}^{D} ( \partial_{\mu} \bd{\Omega})^2 
 + c^{-2} ( \partial_{\tau}\bd{\Omega})^2 \Bigr]
 \nonumber
 \\
& - & \beta V   \frac{ \chi }{2} {h}^2
  +  \frac{  \chi }{2 } \int_0^{\beta} d \tau
 \int d^D r ( \bd{h} \cdot \bd{\Omega} )^2
 \nonumber
 \\
& - & i   \int_0^{\beta} d \tau  
 \int d^D r 
 \bd{M} \cdot ( \bd{\Omega} \times \partial_\tau \bd{\Omega} ) 
\; ,
 \label{eq:sigmah2} 
 \end{eqnarray}
where 
  $\chi  =  {\rho}_s/{ c^2 } $ and
 $\bd{M}  = \chi \bd{h}$.
Choosing the coordinate system such that the staggered magnetization points
in direction $\bd{e}_z$ and keeping in mind that ${\bd{h}} = h {\bd{e}}_x$, we now set 
$\bd{\Omega} = \sqrt{ 1 - \bd{\Pi}^2} {\bd{e}}_z
 + \bd{\Pi}$ and expand Eq.~(\ref{eq:sigmah2}) 
in powers of the transverse fluctuations $\bd{\Pi}.$ Retaining only terms up to cubic order in the
fluctuations $\bd{\Pi} = \Pi_{+} {\bd{e}}_x + \Pi_{-} {\bd{e}}_y$ we obtain in 
momentum-frequency space,
 \begin{eqnarray}
S_{\rm{NLSM} }
[ \bd{\Omega} ] 
&   \approx &  
- \beta V   \frac{ \chi }{2} {h}^2
 \nonumber
\\
& & \hspace{-20mm} +
\frac{ \chi}{2} \int_K \sum_{\sigma}  
\bigl( \omega^2 + c^2 \bd{k}^2  + m^2_{\sigma} \bigr) 
\Pi_{ -K \sigma}  \Pi_{ K  \sigma}
\nonumber
 \\
 &    & \hspace{-20mm} -  
 i   \chi   h   \int_0^{\beta} d \tau \int d^D r  \Pi_{+}^2 
 \partial_\tau \Pi_{ -}
  +  O ( \Pi_{\sigma}^4 ) ,
 \label{eq:Sefffinal1}
 \end{eqnarray}
where $m_{-}^2 =0$ and $m_{+}^2 = h^2$.
At the first sight, the cubic interaction in Eq.~(\ref{eq:Sefffinal1})
does not resemble the cubic term 
$S^{(1/2)}_{\rm eff} [ X_{\sigma} ]$ in Eqs.~(\ref{eq:Seffhalf}--\ref{eq:Gamma3b}).
However, the NLSM is only valid to leading order in the derivatives, so that
for a comparison with Eq.~(\ref{eq:Sefffinal1}) we should expand the
vertices (\ref{eq:Gamma3a}) and (\ref{eq:Gamma3b}) to leading order in 
momenta and frequencies. Moreover, for small $h$ we may approximate
$\Delta_{\bd{k} \sigma} \approx h_c$, so that we obtain
 \begin{eqnarray}
 \Gamma^{(3)}_{ ---} ( K_1 , K_2 , K_3 ) & \approx & \frac{ \lambda}{\sqrt{8S}}
  [     \omega_1   + \omega_2 + \omega_3 ]  =0,
 \nonumber 
\\
 &  & 
\label{eq:gammammm} \\   
 \Gamma^{(3)}_{ -++} ( K_1 ; K_2 , K_3 ) & \approx & \frac{ \lambda}{\sqrt{8S}}
  [     - \omega_1   + \omega_2 + \omega_3 ]
 \nonumber
 \\
 & =  &  - 2 \frac{  \lambda}{\sqrt{8S}}
 \omega_1,
 \end{eqnarray}
where we have used the fact that $\omega_1 + \omega_2 + \omega_3 =0$ by
energy conservation.
Finally, using the relation (\ref{eq:PiX}) between continuum and lattice normalization of the
field representing the staggered spin fluctuations, it is easy to see that
for weak magnetic field
the continuum limit of our  lattice action $S_{\rm eff}^{(1/2)} [ X_{\sigma} ]$
in Eq.~(\ref{eq:Seffhalf}) reduces to the cubic term in the expansion (\ref{eq:Sefffinal1})
of the NLSM.

\section{frequency dependent part of the 
self-energy to order $1/S$}
\label{sec:frequency}

\begin{figure}[tb]
   \centering
   \vspace{7mm}
      \epsfig{file=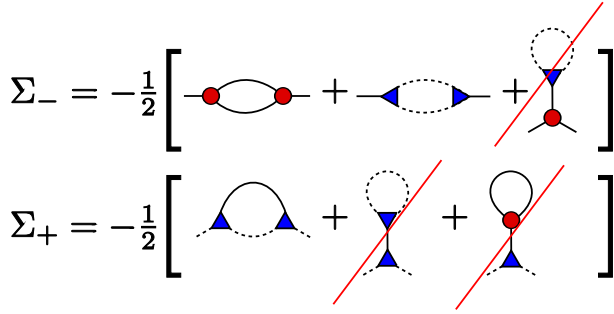,width=80mm}
  \vspace{4mm}
  \caption{%
(Color online) Feynman diagrams of the self-energy corrections to second order
in the three-legged vertices, see Eqs.~(\ref{eq:selfminus}) and (\ref{eq:selfplus}).
The slashed tadpole diagrams give frequency-independent contributions
of order $1/S$ which are analytic functions of the magnetic field.
Since in this work we are only interested in the frequency dependent part of the self-energy,
we shall omit the tadpole diagrams.
}
    \label{fig:self}
  \end{figure}

Defining the non-interacting propagators of the staggered spin fluctuations,
 \begin{equation}
 G_{0, \sigma} (K ) = \frac{ \Delta_{\bd{k} \sigma}}{ E_{\bd{k} \sigma}^2 + \omega^2 },
 \label{eq:G0def}
 \end{equation}
and expressing the corresponding interacting propagators
in terms of the self-energies $\Sigma_{\sigma} (K )$,
   \begin{equation}
 G_{\sigma}^{-1} (  K ) =   G_{0, \sigma}^{-1} (K ) + \Sigma_{\sigma} ( K ),
 \label{eq:Dyson} 
\end{equation}
the leading frequency dependent contribution to the self-energy correction
of the gapless magnon mode can be written as
 \begin{eqnarray}
 \Sigma_{-} ( K ) & = &  \frac{1}{\beta N }   \sum_{ K^{\prime}} \sum_{\sigma}
 G_{0,\sigma} ( K^{\prime} ) G_{0, \sigma} ( K^{\prime} + K )
 \nonumber
 \\
 & & \times
 V^2_{\sigma} (  K , K^{\prime} , -K - K^{\prime} ),
  \label{eq:selfminus}
 \end{eqnarray} 
while the self-energy of the gapped magnon mode is
 \begin{eqnarray}
 \Sigma_{+} ( K ) & = &  \frac{1}{\beta N }   \sum_{ K^{\prime}} 
 G_{0,-} ( K^{\prime} ) G_{0,+} ( K^{\prime} + K )
 \nonumber
 \\
 & & \times
 V^2_{+} (  K^{\prime} , K,  -K - K^{\prime} ),
 \label{eq:selfplus}
 \end{eqnarray} 
where we have used $V_{\sigma} ( -K , -K^{\prime} , K+K^{\prime}  ) = - 
V_{\sigma} (  K , K^{\prime} , -K - K^{\prime} )$.
The corresponding Feynman diagrams are shown in
Fig.~\ref{fig:self}.
The frequency integrations in Eqs.~(\ref{eq:selfminus}) and (\ref{eq:selfplus})
can now be performed analytically; the
relevant integrals are
 \begin{eqnarray}
 I^{(n)} ( E_1 , E_2 , \omega ) & = & \int_{ - \infty}^{\infty} 
 \frac{ d x}{2 \pi} \frac{x^{n}}{ [  x^2 + E_1^2][ ( x + \omega )^2 + E_2^2 ]}
 \nonumber
 \\
 &  & \hspace{-30mm} = \frac{ i^n}{2} \left[
 \frac{ E_1^{n-1}}{E_2^2 - ( E_1 - i \omega)^2} +
 \frac{ ( E_2 + i \omega )^n}{ E_2 [ E_1^2 - ( E_2 + i \omega )^2] }
 \right],
 \end{eqnarray}
where $n=0,1,2$. Explicitly,
 \begin{subequations}
 \begin{eqnarray}
 I^{(0)} & = & \frac{ E_1 + E_2}{2 E_1 E_2 [ (E_1 + E_2)^2 + \omega^2 ] },
 \\
 I^{(1)} & = & - \frac{ \omega}{2  E_2 [ (E_1 + E_2)^2 + \omega^2 ]},
 \\
I^{(2)} & = &  \frac{ E_2 ( E_1 + E_2) + \omega^2 }{2  E_2 [ (E_1 + E_2)^2 + \omega^2 ] }.
\end{eqnarray}
 \end{subequations}
The result for the self-energies can be written as
 \begin{widetext}
\begin{eqnarray}
 \Sigma_{-} ( K ) & = &\frac{h_c^2\lambda ^2}{16S}
\frac 2N\sum_{\bd q}\left\{ z_{\bd q+} z_{\bd k -\bd q+}
 \left[
M_0^2 (\bd k,\bd q)  I^{(0)}_{ ++} (i \omega,\bd k,\bd q)
 +2  M_0 (\bd k,\bd q)   M_+ (\bd k,\bd q)      I^{(1)}_{ ++} (i \omega,\bd k,\bd q)\right.\right.\notag \\
&&\left. \hspace{3.8cm}
+ M^2_{+} (\bd k,\bd q)   I^{(2)}_{++} (i \omega,\bd k,\bd q)
 \right] \notag 
\\
&& \hspace{2cm}+z_{\bd q-} z_{\bd k-\bd q-}
 \left[  M_{-}^2(\bd k,\bd q)  
   I^{(0)}_{--} (i \omega,\bd k,\bd q)  +    2  M_{-} (\bd k,\bd q)  M_{-} (\bd q,\bd k)  
   I^{(1)}_{--} (i \omega,\bd k,\bd q)\notag\right.\\
&&\hspace{3.8cm}\left.\left. +    M_-^2 (\bd q,\bd k)  I^{(2)}_{--} (i \omega,\bd k,\bd q) 
  \right]\right\}, 
 \label{eq:Sigmaminusres}
\\
\Sigma_+(K)&=&
\frac{h_c^2\lambda^2}{16S}\frac 2N \sum_{\bd q} z_{\bd q-} z_{\bd k-\bd q+}
\left[
   M_{+}^2(\bd q,\bd k)    I^{(0)}_{-+} (i \omega,\bd k,\bd q)    
+2    M_+ (\bd q,\bd k)   M_0 (\bd q,\bd k)   I^{(1)}_{-+} (i \omega,\bd k,\bd q)\right.\notag\\
 &&\left.\hspace{3.8cm}+ M_0^2 (\bd q,\bd k)    I^{(2)}_{-+} (i \omega,\bd k,\bd q)
  \right],
 \label{eq:Sigmaplusres}
\end{eqnarray}
\end{widetext}
where
\begin{eqnarray}
I^{(n)}_{ \sigma \sigma^{\prime}} (i \omega,\bd k,\bd q)= \omega^{2-n}
I^{(n)} (E_{\bd q\sigma},E_{\bd k-\bd q\sigma^{\prime}},\omega),
\end{eqnarray}
and we have introduced the functions
\begin{subequations}
\begin{eqnarray}
M_0 (\bd k,\bd q)&=&
\frac{\gamma_{\bd q}}{z_{\bd k-\bd q+}} - 
\frac{\gamma_{\bd k}     -\gamma_{\bd q} -  \gamma_{\bd k-\bd q}   }{z_{\bd k-}},   \\
M_{+} (\bd k,\bd q)&=&\frac{\gamma_{\bd q}}{z_{\bd k-\bd q+}}-\frac{\gamma_{\bd k-\bd q}}{z_{\bd q+}}, \\
M_{-} (\bd k,\bd q)&=&\frac{\gamma_{\bd k}}{z_{\bd k-}}-
\frac{\gamma_{\bd k-\bd q}}{z_{\bd k-\bd q-}}.
\end{eqnarray}
\end{subequations}
For later reference we note that
 \begin{subequations}
 \begin{eqnarray}
   M_0 ( 0 , \bd{q} ) & = &  \frac{ \gamma_{\bd{q}} }{ z_{\bd{q}+}} 
 + \frac{ 2 \gamma_{\bd{q}} -1 }{z_{0-}},
 \label{eq:A1}
 \\
 M_+ ( 0 , \bd{q} ) & = & 0,
 \\
   M_- ( 0 , \bd{q} ) & = & \frac{1}{z_{0-}} - \frac{  \gamma_{\bd{q}}}{ z_{\bd{q} -} }   ,
 \\
M_0 (  \bd{k} ,0 ) & = &  \frac{1}{ z_{\bd{k}+}} +   \frac{1}{ z_{\bd{k}-}},
 \\
  M_+ (  \bd{k},0 ) & = & \frac{1}{z_{ \bd{k} +}} - \frac{  \gamma_{\bd{k}}}{ z_{0 +} }   ,
  \\
M_- (  \bd{k} ,0) & = & 0.
 \label{eq:A6}
 \end{eqnarray}
\end{subequations}
Furthermore, if both $\bd{k}$ and $\bd{q}$ are small
 \begin{equation}
 M_{-} ( \bd{k} , \bd{q} ) = \frac{a^2}{4 D n^4} \left[ 
 \bd{q}^2 - 2 \bd{k} \cdot \bd{q} \right] + O ( k^4, q^4, k^2 q^2 ).
 \label{eq:Aminussmall} 
\end{equation}

\section{Renormalization of the gapless magnon}
\label{sec:renormalization}

\subsection{Spin-wave velocity}
\label{subsec:velocity}

We now show that in dimensions $D \leq 3$ the leading $1/S$-correction to the
spin-wave velocity $\tilde{c}_- ( h )$ of the gapless magnon is non-analytic in $h^2$. 
Therefore we expand for small  $\omega$ and $ | \bd{k}|$,
  \begin{eqnarray}
 \Sigma_{-} ( \bd{k} , i \omega ) & = &  
 f_{0} \omega^2 + f_{1}  \omega_{\bd{k}}^2  
 \nonumber
 \\   
& + &  f_2  \omega^4 + f_3 \omega^2 \omega_{\bd{k}}^2  +  
f_{4}   \omega_{\bd{k}}^4   + O ( \omega^6 ), 
 \label{eq:Sigmaplexpan}
 \end{eqnarray}
where $\omega$ and $\omega_{\bd{k}} = c_- | \bd{k} |$ 
are assumed to have the same order of magnitude and
$c_{ -} = c_0 n^2 $ is the spin-wave velocity within linear spin-wave theory,
see Eq.~(\ref{eq:cminusnull}).   To calculate the renormalized spin-wave velocity
we may neglect in Eq.~(\ref{eq:Sigmaplexpan})  the terms of order $\omega^4$  involving the coefficient $f_2$, $f_3$ and $f_4$.
Using Eqs.~(\ref{eq:G0def}) and (\ref{eq:Dyson}) we  obtain for the
infrared behavior of the
propagator of the gapless mode 
 \begin{equation}
 G_{-} ( \bd{k} , i \omega ) =  \frac{ Z_- h_c n^2  }{ \omega^2 +  \tilde{c}_-^2 \bd{k}^2 }.
 \label{eq:Gminus1}
 \end{equation}
Introducing the dimensionless constants ${F}_{0}$ and ${F}_{1}$,
 \begin{equation}
 {F}_{0} = h_c n^2 f_{0} , \; \; \;  {F}_{1} = h_c n^2 f_{1},
 \end{equation}
the wave-function renormalization factor $Z_-$ can be written as
 \begin{equation}
 Z_- = \frac{1}{ 1 +  {F}_{0} } \approx 1 - F_{0},
 \end{equation}
and the renormalized spin-wave velocity $\tilde{c}_-$ obeys
 \begin{equation}
 \frac{\tilde{c}_-^2}{  c_-^2} =  \frac{ 1 + {F}_{1}}{1 +  {F}_{0}} \approx 1 + F_{1} - F_{0}.
 \label{eq:cminusratio}
 \end{equation}
The constants $f_{0}$ and $f_2$ associated with
the expansion in powers of frequencies for vanishing external momentum
can be obtained by expanding
$\Sigma_- ( \bd{k} =0, \omega )$ in powers of $\omega^2$. Using Eq.~(\ref{eq:Sigmaminusres}) and
Eqs.~(\ref{eq:A1}--\ref{eq:A6}) one gets
 \begin{eqnarray}
 \Sigma_- ( 0, i \omega ) & = & 
 \frac{h_c^2\lambda ^2}{16S}
\frac 2N\sum_{\bd q} \Bigl\{ z_{\bd q+}^2
M_0^2 (0,\bd q)  I^{(0)}_{ ++} (i \omega,0,\bd q)
 \nonumber
 \\
&  &  \hspace{18mm} + z_{\bd q-}^2
M_-^2 (0,\bd q)  I^{(0)}_{ --} (i \omega,0,\bd q)
\Bigr\}
 \nonumber
 \\
 & = &  \omega^2 
\frac{h_c^2\lambda ^2}{16S}
\frac 2N\sum_{\bd q} \biggl\{
\frac{ \bigl[ \gamma_{\bd{q}} + \frac{ 1 + \gamma_{\bd{q}} }{2 n^2} 
( 2 \gamma_{\bd{q}} -1 ) \bigr]^2}{ E_{\bd{q} +}  [ (2 E_{\bd{q} +})^2 + \omega^2 ] }
\nonumber
 \\
 & & \hspace{15mm} + \frac{ \bigl[ \frac{ z_{\bd{q} -}}{z_{0-}} - 
\gamma_{\bd{q}} \bigr]^2}{ 
 E_{\bd{q} -}  [ (2 E_{\bd{q} -})^2 + \omega^2 ] }
\biggr\}.
 \end{eqnarray}
Using $ h_c^2 \lambda^2 = 4 n^2 h^2$, we obtain for the
first two coefficients in the frequency expansion,
 \begin{eqnarray}
 f_{0} & = & 
\frac{n^2 h^2}{16S}
\frac 2N\sum_{\bd q} \biggl\{
\frac{ \bigl[ \gamma_{\bd{q}} + \frac{ 1 + \gamma_{\bd{q}} }{2 n^2} 
( 2 \gamma_{\bd{q}} -1 ) \bigr]^2}{  E_{\bd{q} +}^3   }
\nonumber
 \\
 & &  \hspace{20mm} + \frac{  
\bigl[ \frac{ z_{\bd{q} -}}{z_{0-}} - \gamma_{\bd{q}} \bigr]^2    }{ 
  E_{\bd{q} -}^3  } \biggr\},
 \label{eq:g20}
 \end{eqnarray}
\begin{eqnarray}
 f_{2} & = & 
- \frac{n^2 h^2}{16S}
\frac 2N \sum_{\bd q} \biggl\{
\frac{ \bigl[ \gamma_{\bd{q}} + \frac{ 1 + \gamma_{\bd{q}} }{2 n^2} 
( 2 \gamma_{\bd{q}} -1 ) \bigr]^2}{  4 E_{\bd{q} +}^5   }
\nonumber
 \\
 & &  \hspace{20mm} + \frac{ \bigl[ 
\frac{ z_{\bd{q} -}}{z_{0-}} - \gamma_{\bd{q}} \bigr]^2     }{ 
  4 E_{\bd{q} -}^5  } \biggr\}.
 \label{eq:g40}
 \end{eqnarray}
Keeping in mind that $  z_{\bd{q} -}/{z_{0-}}  - \gamma_{\bd{q}} = O ( q^2)$ for small $q$,
it is easy to see that in the domain of
small magnetic field $( h \ll h_c)$
the integrals on the right-hand sides
of the equations above
are dominated by the first term involving
the gapped mode $E_{\bd{q} +}$.
More precisely, the relevant ultraviolet cutoff
for the momentum integrals in Eqs.~(\ref{eq:g20}, \ref{eq:g40}) 
is the inverse of the length scale
 \begin{equation}
 \xi =   c_0/h .
 \label{eq:xidef}
 \end{equation}
In $D \leq 3$ the contribution from wave-vectors
in the regime $ | \bd{q} | \xi \lesssim 1$
gives rise to contributions to the magnon self-energy 
which are non-analytic in $h^2$.
Keeping in mind that for small field the magnetic length $\xi$ is large compared with the
lattice spacing,
we may calculate the leading non-analytic magnetic-field dependent contributions
to Eqs.~(\ref{eq:g20}, \ref{eq:g40}) by expanding 
the integrand in powers of $\bd{q}$. 

We find that the leading magnetic field dependence of the spin-wave velocity
$\tilde{c}_-$ associated with the gapless mode is determined by $f_{0}$.
Since we are only interested in the non-analytic $h^2$-dependence,
we may set $n \approx 1$. In the thermodynamic limit 
we then obtain for the dominant contribution to Eq.~(\ref{eq:g20}),
 \begin{eqnarray}
 f_{0} & \approx & \frac{h^2 a^D}{2S} \int \frac{ d^D q}{ (2 \pi )^D}
 \frac{1}{ E_{\bd{q} +}^3}.
 \end{eqnarray}  
Consistently neglecting terms which are analytic in $h^2$, 
we may ignore the magnetic field dependence of the non-interacting spin-wave velocities,
$c_{\pm} \approx c_0  = 2 \sqrt{D} JS a$, so that energy dispersions are approximated by
$E_{\bd{q} - } \approx c_0 | {\bd{q}} |$ and
 $E_{\bd{q} + } \approx \sqrt{ h^2 + c^2_0  {\bd{q}}^2}$.
Using  $h_c = 2 \sqrt{D} c_0/a$ we obtain 
from Eq.~(\ref{eq:g20}) for the corresponding dimensionless coefficient
for $1 < D \leq 3$,
 \begin{equation}
 {F}_{0} = h_c n^2 f_{0} =   \alpha_D  \frac{ m^{D-1}   }{S},
 \label{eq:gomegaes}
 \end{equation}
where $m = h / h_c = h a /(2 \sqrt{D} c_0)$
is the relevant dimensionless magnetic field
[see Eq.~(\ref{eq:cantingclassical})], and
 \begin{equation}
 \alpha_D =    2^{D-1} D^{D/2} K_D \int_0^{1/m} dy \frac{ y^{D-1}}{[ 1 + y^2]^{3/2} }.
 \end{equation}
Here 
 \begin{equation}
 K_D = \frac{2^{1-D}}{\pi^{D/2} \Gamma ( D/2)}
 \end{equation}
is the surface area of the $D$-dimensional unit sphere divided by
$(2 \pi )^D$.
In $D < 3$ we may take the limit $1/m \rightarrow \infty$ in $\alpha_D$, so that
\begin{equation}
 \alpha_D  =  \left( \frac{D}{\pi} \right)^{D/2}
 \frac{\Gamma ( \frac{3-D}{2} ) }{ \sqrt{\pi}}.
 \end{equation}
In particular, $\alpha_2 = 2/\pi$. 
In $D=3$ the integral $\alpha_3$ depends for small $m$  
logarithmically on the upper limit,
 \begin{equation}
 \alpha_3 \sim  \alpha_3^{\prime}  \ln (1/ m) , \; \; \alpha_3^{\prime} =
\frac{6 \sqrt{3}}{\pi^2}.
 \end{equation}
It turns out that the coefficient $F_{1}$ in front of the
$\bd{k}^2$-correction to the self-energy is for small $h$
proportional to $h^2$, so that for $h \ll \Delta$ the dominant 
magnetic-field dependence of the
spin-wave velocity is due to the term $ F_{0}  $
in Eq.~(\ref{eq:cminusratio}). We thus obtain for the leading
magnetic field dependence of the spin-wave velocity of the gapless magnon
 \begin{subequations}
 \begin{eqnarray}
 \frac{\tilde{c}_-^2 }{c_{0}^2}   & \approx  & 1 - {F}_{0}
 \nonumber
 \\
  & = & 1   - \frac{2}{\pi S}  \frac{ | \bd{h} | }{h_c}, \; \;  D=2, 
 \label{eq:cres1} 
\\
& = & 1   -
  \frac{6 \sqrt{3}}{  \pi^{2} S }   \frac{ \bd{h}^2}{h_c^2} 
\ln \left( \frac{h_c }{ | \bd{h} | } \right),  \; \;  D=3,
 \label{eq:cres}
 \end{eqnarray}
\end{subequations}
where we have neglected  magnetic field independent $1/S$-corrections.
Recall that within linear spin-wave theory the 
velocity $c_-$ of the gapless magnon is
analytic in $\bd{h}^2 = h_c^2 m^2$; 
from  Eq.~(\ref{eq:cminusnull}) we obtain 
$c_- \approx c_0 [ 1 - m^2/2 ]$ for small $m$.
We conclude that in dimensions $D \leq 3$  
the dominant  magnetic field dependence of the spin-wave velocity
of the gapless magnon
is due to spin-wave interactions.
In Appendix~B we show that the
non-analytic dependence on $\bd{h}^2$
predicted by  Eq.~(\ref{eq:cres1}) can be recovered numerically from 
in the expression for the magnon self-energy given by
Zhitomirsky and Chernyshev \cite{Zhitomirsky99}.

\subsection{Quasiparticle residue}
\label{subsec:cubic}

In view of the fact that the magnetic field dependence of the
spin-wave velocity of the gapless magnon is
dominated by spin-wave  interactions, it is reasonable to   expect that
also the higher coefficients in the expansion of the self-energy
of the gapless magnon
for small wave-vectors and frequencies exhibit some non-analytic dependence
on the magnetic field. Consider first the renormalized magnon energies  $\tilde{E}_{\bd{k} \sigma}$, which can be defined by
\begin{equation}
 \tilde{E}_{\bd{k} \sigma}^2 = E_{\bd{k} \sigma}^2 + \Delta_{\bd{k} \sigma} 
 {\rm Re} \Sigma_\sigma ( \bd{k} , \tilde{E}_{\bd{k} \sigma} + i 0 ) .
 \end{equation}
The expansion for small wave-vectors is
\begin{equation}
 \tilde{E}^2_{\bd{k}-}  =  \tilde{c}_{-}^2 \bd{k}^2  \left[ 1 +  \tilde{A}_{-}  ( \hat{\bd{k}} )    
\bd{k}^2   + O ( \bd{k}^4 ) \right] .
 \label{eq:dispren}
 \end{equation}
It is well known \cite{Landau80} that only if the  coefficient $\tilde{A}_{-} $ is  positive
a  gapless magnon with momentum
$\bd{k}$ can spontaneously decay
into two  magnons with momenta $\bd{q}$  and $ {\bd{k}} - \bd{q}$.
Within linear spin-wave theory we obtain from Eqs.~(\ref{eq:epsilondef}) and
(\ref{eq:Edispersiondef}) in $D$ dimensions
\begin{eqnarray}
 {E}^2_{\bd{k}-} & = & c_-^2 \bd{k}^2  \left[ 1 +  A_{-}  ( \hat{\bd{k}} )    
\bd{k}^2   + O ( \bd{k}^4 ) \right] ,
 \label{eq:Erenexpansion}
 \\
 {E}^2_{\bd{k}+} & = & h^2 +  c_+^2 \bd{k}^2  
\left[ 1 +  A_{+} ( \hat{\bd{k}} )     \bd{k}^2   + O (  \bd{k}^4 ) \right] ,
 \hspace{7mm}
 \label{eq:Erenexpansionplus}
 \end{eqnarray}
with
 \begin{eqnarray}
 A_{-} ( \hat{\bd{k}} ) & = &  - 
 \frac{a^2}{4} \left[ \frac{1 - 2 m^2 }{D (1-m^2) }    +    \frac{1}{3} \sum_{\mu}   
\hat{k}_{\mu}^4 \right],
 \label{eq:Aminus}
 \\
A_{+}( \hat{\bd{k}} ) & = &  - 
 \frac{a^2}{4} \left[ \frac{1 - 2 m^2 }{D (1-3m^2) }    +    \frac{1}{3} \sum_{\mu}   
\hat{k}_{\mu}^4 \right].
  \label{eq:Aplus}
 \end{eqnarray}
Obviously, for $ m \ll  1$ the coefficient $A_{-} ( \hat{\bd{k}} )$ 
is negative for all directions $\hat{\bd{k}}$,
so that to this order in spin-wave theory the gapless
magnon cannot spontaneously decay at long wave-lengths.
For larger $m$
the coefficient $A_{-} ( \hat{\bd{k}} )$ decreases and
eventually vanishes at a critical $m_{\ast} ( \hat{\bd{k}} )$ which 
depends on the direction $\hat{\bd{k}}$.
From  Eq.~(\ref{eq:Aminus}) it is easy to show  that the direction where 
$m_{\ast} ( \hat{\bd{k}} )$ assumes the smallest possible 
value is given by the diagonal $\hat{k}_x =  \ldots = \hat{k}_D$, and that the associated
minimum is $m_{\ast} = h_{\ast} / h_c = 2 / \sqrt{7} \approx 0.76$.
For the special case $D=2$ this result has been obtained previously by
Zhitomirsky and Chernyshev \cite{Zhitomirsky99}, who examined the 
leading $1/S$-correction 
in the regime $ h_{\ast} < h < h_c$ numerically.

Apparently, the leading $1/S$-correction in the limit of small magnetic fields
$ m = h/ h_c \ll 1$ has not been explicitly analyzed
in Ref.~[\onlinecite{Zhitomirsky99}].
In terms of the expansion coefficients introduced in Eq.~(\ref{eq:Sigmaplexpan})
we obtain 
$\tilde{A}_-  ( \hat{\bd{k}} ) = A_- ( \hat{\bd{k}} )   + \delta A_-$, where 
the $1/S$-correction is
 \begin{equation}
 \delta A_- =    c_0^2 h_c \left[ f_{2} - f_{3} + f_{4} \right].
 \label{eq:Ami}
 \end{equation}
Let us consider first the contribution from the coefficient $f_2$ related to the
  $\omega^4$-term 
in the expansion of the self-energy $ \Sigma_- (0 , i \omega )$ for small 
frequencies. Because for small $h$ the integral defining
$f_2$ in Eq.~(\ref{eq:g40}) the dominated by wave-vectors $ |\bd{q} | \lesssim h / c_0$,
we may approximate
 \begin{equation}
f_{2}  \approx  - \frac{h^2 a^D}{8S} \int \frac{ d^D q}{ (2 \pi )^D}
 \frac{1}{  E_{\bd{q} +}^5}.
 \end{equation}
The integral is easily evaluated to leading order for small $m \ll 1$.
Introducing the dimensionless coefficient
 \begin{equation}
 F_2 \equiv \frac{ c_0^2 h_c f_2}{a^2},
 \label{eq:F2def}
 \end{equation}
we obtain for $ D < 3$,
 \begin{equation}
 F_2 \approx - \frac{ \beta_D }{S} \left[  m^{D-3} + O ( m^{D-1}  ) \right] ,
 \label{eq:F2res}
 \end{equation}
with the numerical coefficient
 \begin{eqnarray}
 \beta_D & = &  \frac{( 2 \sqrt{D} )^{D-2}}{8}
 K_D  \int_0^{\infty} dx \frac{x^{D-1}}{ [ 1
 + x^2]^{5/2} }
 \nonumber
 \\
& = & \frac{( 2 \sqrt{D} )^{D-2}}{8}
 K_D \frac{2}{3 \sqrt{\pi}} \Gamma ( \frac{5-D}{2} ) \Gamma ( \frac{D}{2} ).
 \label{eq:betaDdef}
 \end{eqnarray}
In particular, in two dimensions $\beta_2 = 1/(48 \pi)$.
Obviously, for $D < 3$ the coefficient $F_2$ diverges for $m \rightarrow 0$,
so that the contribution from  the term $f_2$ to 
$\delta A_-$  is for sufficiently small $m$  much larger than the linear spin-wave
result (\ref{eq:Aminus}).
It turns out, however, that  the singular contribution   to
$\delta A_-$ due to $f_2$ is exactly canceled
by a similar contribution from the  coefficient $f_3$.
In order to extract the dominated contribution to
$f_3$, it is sufficient
to approximate the magnon self-energy (\ref{eq:Sigmaminusres}) by
 \begin{equation}
 \Sigma_{-} ( \bd{k} , i \omega ) \approx \frac{2 h^2 a^D}{S}
 \int \frac{ d^D q}{(2 \pi )^D} I_{++}^{(0)} ( i \omega , {\bd{k}} , {\bd{q}} ).
 \label{eq:sigmasing} 
\end{equation}
Expanding the right-hand side to second order in $\bd{k}$ and comparing with Eq. (\ref{eq:Sigmaplexpan}), we obtain
 \begin{eqnarray} 
 f_{3} & \approx & - \frac{h^2 a^D}{8S} \int \frac{ d^D q}{ (2 \pi )^D}
 \frac{1}{  E_{\bd{q} +}^5} \left[ 3 - \frac{10}{D} \frac{ c_0^2 \bd{q}^2}{ E_{\bd{q}+}^2}
 \right].
 \end{eqnarray}
The integral can  easily be carried out analytically with the result $f_3 = f_2 + O ( m^{D-1 })$.
From Eq.~(\ref{eq:Sigmaminusres}) we can also show that
the term $f_4$ is of order  $a^2 m^{D-1}/S$ and can be neglected as compared with
$f_2$ and $f_3$.
Because $\delta A_-$ involves the combination $f_2 - f_3$,  we conclude that the singular contributions proportional to $m^{D-3}$
cancel in $\delta A_-$, so that the leading magnetic field dependence of
$A_-$ is proportional to $m^{D-1} \propto | \bd{h} |^{D-1}$.
This  is small
compared with the linear spin-wave result
but non-analytic in $\bd{h}^2$, similar to the leading magnetic field-dependence
of the spin-wave velocity in Eqs.~(\ref{eq:cres1}, \ref{eq:cres}).

On the other hand, the singular magnetic field dependence appearing
in the coefficients
 $f_2$ and $f_3$ does not cancel in the self-energy 
$\Sigma_- ( \bd{k} , \omega + i 0 )$ off resonance. Retaining only
the singular contributions to Eq.~(\ref{eq:Sigmaminusres}) we obtain
with $f_2 \approx f_3$
 \begin{equation}
 \Sigma_- ( \bd{k} , \omega + i 0 ) \approx
 - f_0 \omega^2  + f_2 \omega^2 ( \omega^2 - \tilde{c}_-^2 \bd{k}^2 ).
 \end{equation}
The corresponding renormalized magnon Green function for small
$m$ can be written as
  \begin{equation}
 G_{-} ( \bd{k} , i \omega  ) = Z_- ( i \omega ) \frac{  h_c n^2  }{  
\omega^2 + \tilde{c}_-^2 \bd{k}^2    },
 \label{eq:Gminus2}
 \end{equation}
 where the renormalized spin-wave velocity is given in
Eqs.~(\ref{eq:cres1},\ref{eq:cres}), and 
 \begin{eqnarray}
 Z_- ( i \omega ) & = & \frac{ 1}{ 1 + F_0  +  h_c n^2 f_2 \omega^2 }
 \nonumber
 \\
 & \approx & 1 - F_0 -  ( a^2 F_2/ c_0^2)  \omega^2.
 \end{eqnarray}
After analytic continuation to real frequencies we obtain for the
renormalized residue of the magnon peak for small $m$,
 \begin{eqnarray}
 Z_{\bd{k} -} \equiv Z_- ( i \omega \rightarrow \tilde{c}_-  | \bd{k} | ) & = & 1 - F_0 + F_2   \bd{k}^2 a^2
 \nonumber
 \\
 & = &  1 - F_0 -  \beta_D   \frac{  \bd{k}^2 a^2 }{S m^{3-D}}. \hspace{10mm}
 \end{eqnarray}
Expressing
$m = h / h_c = h a /( 2 \sqrt{D} c_0)   = a /( 2 \sqrt{D} \xi ) $ in terms of the length scale
$\xi = c_0 / h$  associated with the magnetic field 
we may alternatively write
  \begin{eqnarray}
   Z_{\bd{k} -}  & = & 1 - F_0 - \frac{\tilde{\beta}_D}{S}  \Bigl( \frac{\xi }{ a } \Bigr)^{3-D}  
 \bd{k}^2 a^2
 \nonumber
 \\
 & = &  1 -\frac{1}{S} \Bigl( \frac{a}{  \xi} \Bigr)^{D-1} \left[\tilde{\alpha}_D - \tilde{\beta}_D     \bd{k}^2 \xi^2\right],
 \label{eq:Zkminus}
 \end{eqnarray}
where $\tilde{\alpha}_D=\alpha_D (2\sqrt{D})^{1-D}$ and  $\tilde{\beta}_D = \beta_D ( 2 \sqrt{D} )^{3-D}$.
In particular, in $D=2$ the leading momentum dependence
of $Z_{\bd{k} -}$ is proportional to  $ \bd{k}^2 \xi a =  \bd{k}^2  c_0 a/ h$.
The higher powers in $\bd{k}$  become important for
 $ | \bd{k} \xi | \gtrsim  1$, so that the expansion 
(\ref{eq:Zkminus}) is limited to the regime $ | \bd{k} | \lesssim \xi^{-1} \ll a^{-1}$
where the $1/S$-correction is small compared with unity.

\subsection{Magnon damping}

Given the magnon self-energies $\Sigma_{\sigma} ( K )$ in
Eqs.~(\ref{eq:Sigmaminusres},\ref{eq:Sigmaplusres})
and the renormalized magnon dispersions $\tilde{E}_{\bd{k} \sigma}$,
the magnon damping can be obtained from
 \begin{equation}
 \Gamma_{\bd{k} \sigma} = - \frac{ \Delta_{\bd{k} \sigma}}{ 2 \tilde{E}_{\bd{k} \sigma}}
 {\rm Im} \Sigma_{\sigma}  ( \bd{k} , \tilde{E}_{\bd{k} \sigma} + i 0 ).
 \label{eq:Imdef} 
\end{equation} 
Zhitomirsky and Chernyshev \cite{Zhitomirsky99} have shown that
in two dimensions one should 
self-consistently take into account the imaginary
part of the magnon self-energy when evaluating
the integrals on the right-hand side of Eq.~(\ref{eq:Sigmaminusres}).
However, as long as we are not too close to the critical field $h_*$,
the result for the magnon damping is
non-singular even if we ignore the
damping of intermediate magnons
in Eq.~(\ref{eq:Sigmaminusres}).
We therefore
expect that a simplified version
of Eq.~(\ref{eq:Imdef}) taking into account only the renormalization of the
real part of the magnon dispersion
yields a qualitatively correct estimate for the magnon damping away from $h_*$.

To calculate the damping $\Gamma_{\bd{k} - }$ of the gapless magnon for
wave-vectors $ | \bd{k} | \ll h/ c_0 = \xi^{-1}$, it is sufficient
to retain in Eq.~(\ref{eq:Sigmaminusres}) only the terms
involving the functions $ I^{(n)}_{ --} ( i \omega , \bd{k} , \bd{q} )$, because
the imaginary part of the functions $ I^{(n)}_{ ++} ( \omega + i 0 , \bd{k} , \bd{q} )$ vanishes
for $ \omega < 2 h$.
Using Eq.~(\ref{eq:Aminussmall}) we obtain for $\omega >0$
 \begin{widetext} 
\begin{eqnarray}
 & &
 {\rm Im} \left[
 M_{-}^2(\bd k,\bd q)  
   I^{(0)}_{--} ( \omega +  i 0 ,\bd k,\bd q)  +    2  M_{-} (\bd k,\bd q)  M_{-} (\bd q,\bd k)  
   I^{(1)}_{--} ( \omega + i 0 ,\bd k,\bd q)
+    M_-^2 (\bd q,\bd k)  I^{(2)}_{--} ( \omega + i 0 ,\bd k,\bd q)  \right] 
 \nonumber
 \\
 & & = -
 \frac{\pi}{4}   \left( \frac{ a^2}{ 4 D n^4 } \right)^2
 W ( \bd{k} , \bd{q} )
 \delta (  \omega  - \tilde{E}_{\bd{k} - \bd{q}- } - \tilde{E}_{\bd{q} -} ),
 \label{eq:ImM}
\end{eqnarray}
 \end{widetext}
where
 \begin{eqnarray}
 W ( \bd{k} , \bd{q} ) & = & \frac{ q}{ | \bd{k} - \bd{q} | } ( k^2 - q^2 )^2 +
\frac{ | \bd{k} - \bd{q} | }{q} ( q^2 - 2 \bd{k} \cdot \bd{q} )^2 
 \nonumber
 \\
&  & - 2 ( k^2 - q^2)
 ( q^2 - 2 \bd{k} \cdot \bd{q} ).
 \label{eq:Wdef}
 \end{eqnarray}
Note that in the non-linear sigma model the contribution corresponding
to Eq.~(\ref{eq:ImM}) is neglected because  the relevant vertex
 involving three gapless magnons is set equal to zero (see Eq. (\ref{eq:gammammm})), which is
correct to leading order in the derivatives. 
Hence, the damping of the gapless magnon cannot be
obtained using the  NLSM.
To estimate the magnon damping we set
$ \omega = \tilde{E}_{\bd{k} - }  $
and approximate the renormalized magnon dispersion by
 \begin{equation}
 \tilde{E}_{\bd{k} - } \approx
 c_-  | \bd{k} | ( 1 + \bar{A}_- \bd{k}^2 ),
 \end{equation} 
where for simplicity we have replaced the direction-dependent coefficient
$\tilde{A}_- ( \hat{\bd{k}} )$ defined in
Eq.~(\ref{eq:dispren}) by some angular average $\bar{A}_-$.
At long wave-lengths we then obtain
 \begin{eqnarray}
 \Gamma_{\bd{k} -} & = & \frac{ \pi \sqrt{D}}{ 8 (4 D)^2 } \frac{h^2 a^{D+3}}{S  }
 \int \frac{ d^D q}{ (2 \pi )^D}  \frac{W ( \bd{k} , \bd{q} )}{k} 
 \nonumber
 \\
 & & \times
 \delta (  \tilde{E}_{\bd{k}-}  - \tilde{E}_{\bd{k} - \bd{q}- } - \tilde{E}_{\bd{q} -} ).
 \label{eq:Gammak1}
 \end{eqnarray}
As discussed in the textbook by Lifshitz and Pitaevskii \cite{Landau80},
in the long wave-length limit  
the  energy conservation
 $\tilde{E}_{\bd{k} - } =
  \tilde{E}_{\bd{k} - \bd{q}- } + \tilde{E}_{\bd{q} -}$ 
can only be satisfied for $\bar{A}_- > 0$.
From our discussion in Sec.~\ref{subsec:cubic}  
(see also Ref.~[\onlinecite{Zhitomirsky99}]) we know that this condition is only satisfied 
in a certain range  $ h_{\ast} < |\bd{h}| < h_c$ of magnetic fields
below the saturation field.
We now restrict ourselves to this regime, without explicitly calculating
the magnetic-field dependence of the coefficient $\bar{A}_-  >0 $.
If $h$ is not very close to the threshold fields $h_{\ast}$ and $h_c$, 
we expect by dimensional analysis that  $\bar{A}_- / a^2$ 
is a number of the order of unity.
The energy conservation
then implies that the allowed vectors $\bd{q}$ are almost parallel to
the direction of $\bd{k}$ and satisfy $q \leq k$. In fact, it is easy to show that
the angle $\vartheta$ between $ \bd{k}$ and $\bd{q}$ 
is  $ \vartheta \approx \sqrt{6 \bar{A}_- } ( k-q )$ due to energy conservation so that for $\bar{A}_- k^2 \ll 1$
we may approximate
 \begin{equation}
 \delta ( \tilde{E}_{\bd{k} -}   - \tilde{E}_{\bd{k} - \bd{q}- } - \tilde{E}_{\bd{q} -} )
 \approx \frac{ \delta ( \vartheta -  \sqrt{6 \bar{A}_- } ( k-q ) ) }{ \sqrt{6 \bar{A}_- } c_- k q },
 \end{equation}
and
 \begin{eqnarray}
 | \bd{k} - \bd{q} | & \approx &  k -q + \frac{kq}{k-q} ( 1 - \cos \vartheta ) 
 \nonumber
 \\
 & \approx &  ( k-q ) ( 1 + 3 \bar{A}_- k q ).
 \end{eqnarray}
Keeping in mind that $ \bar{A}_- k q \ll 1$ we obtain from Eq.~(\ref{eq:Wdef}), 
 \begin{eqnarray}
 \frac{W ( \bd{k} , \bd{q} )}{k} 
 \approx 9 k q ( k-q ) .
 \end{eqnarray}
The integrations in Eq.~(\ref{eq:Gammak1}) are now elementary 
and we obtain for the  damping of
magnons with  wave-vectors in the regime
$| \bd{k} | \lesssim  h / c_0 \ll a^{-1}$ at zero temperature  in $D$ dimensions,
 \begin{equation}
 \frac{ \Gamma_{\bd{k} -}}{ E_{\bd{k} -}  } = \frac{ \gamma_D}{ S}
  \left(  \frac{h }{ h_c} \right)^2 \left( \sqrt{ 6 \bar{A}_- } \right)^{D-3} a^{D+1}
   | \bd{k}|^{2D-2}   ,
 \label{eq:decayres}
 \end{equation}
where 
  \begin{eqnarray}
 \gamma_D & = & \frac{9   }{64 \sqrt{D}}    K_{D-1}
 \int_0^1 dx [ x (1-x)]^{D-1}
 \nonumber
 \\
 & = &   \frac{9    }{64 \sqrt{D}}  
 K_{D-1}
 2^{1-2D}  \frac{ \sqrt{\pi}  \Gamma ( D )}{\Gamma( D + \frac{1}{2} ) }.
 \end{eqnarray}
In two dimensions we have $\gamma_2 = 3  /(128 \sqrt{2}  \pi )$ and
 \begin{equation}
 \Gamma_{\bd{k} -} = \frac{ \gamma_2}{ S  }
 \left(  \frac{h }{ h_c}  \right)^{2}  \frac{ \tilde{c}_- | \bd{k} |^3 a^3   }{\sqrt{6 \bar{A}_-}}   .
 \end{equation}
The $| \bd{k} |^3$-dependence of the magnon damping has been obtained previously by
Zhitomirsky and Chernyshev \cite{Zhitomirsky99}.

\section{Summary and Conclusions}
\label{sec:conclusions}
The main result of this work is the discovery that 
in quantum Heisenberg antiferromagnets subject to
a weak uniform external field the 
leading $1/S$-correction to the self-energy of the  gapless magnon
 is a non-analytic function of $\bd{h}^2$ in dimensions $ D \leq 3$.
We have explicitly calculated the leading magnetic field
dependence of the
spin-wave velocity and the 
momentum-dependent quasi-particle residue of the gapless magnon.
At the first sight it is surprising that  for quantum antiferromagnets
in a uniform magnetic field at zero temperature the dimension
 $D=3$ plays the role of a critical dimension below which
fluctuations lead to a non-analytic magnetic field dependence 
of the magnon spectrum.
However,  the gapless magnons  in our model
can be viewed as  an interacting Bose gas in the condensed phase \cite{Kreisel07}, 
where the Bogoliubov mean-field theory 
is known \cite{Castellani97,Wetterich07} to break down
in dimensions $D \leq 3$.

Finally, let us point out that our  hybrid approach between
$1/S$-expansion and NLSM
is  a very convenient   parameterization 
of the spin-wave expansion, which should also be useful in other contexts.
While the calculations presented here can (with some effort)  also be carried out
using the conventional parameterization of the $1/S$-expansion,
our hybrid approach  greatly facilitates  the identification of the
frequency dependent contributions to the magnon self-energies 
which give  rise to the dominant magnetic field dependent corrections 
to the magnon spectrum.

\section*{ACKNOWLEDGMENTS}
We thank A.~L.~Chernyshev and M.~E.~Zhitomirsky for interesting discussions.
This work was financially supported by the DFG via SFB/TRR 49, FOR 412, and
by the DAAD via the PROBRAL program.

\appendix

\renewcommand{\theequation}{A.\arabic{equation}}
\renewcommand{\thesubsection}{\arabic{subsection}}
 \begin{widetext}
\section{Quartic spin-wave interaction in Hermitian field 
parameterization}

In Hermitian field parameterization, the quartic part of the
Hamiltonian $\hat{H}_4^{\prime}$ defined in Eqs.~(\ref{eq:H4},\ref{eq:H4sum}) is
{\small
\begin{eqnarray}
 \hat{H}^{\prime}_4 & = &
     \frac{ 2}{N} \sum_{ \bd{k}_1  \bd{k}_2  \bd{k}_3 \bd{k}_4 }
 \delta_{ \bd{k}_1 + \bd{k}_2 + \bd{k}_3 + \bd{k}_4 , 0}
\nonumber
\\
 & \times & 
\biggl[  \frac{1}{4!} 
 \Bigl( \Gamma^{ PPPP}_{++++} ( \bd{k}_1 , \bd{k}_2 , \bd{k}_3 , \bd{k}_4 )
 \hat{P}_{ \bd{k}_1 +}  \hat{P}_{ \bd{k}_2 +}  \hat{P}_{ \bd{k}_3 +}  \hat{P}_{ \bd{k}_4 +} 
%\nonumber
% \\
% & & \hspace{39mm} 
   + 
  \Gamma^{ PPPP}_{----} ( \bd{k}_1 , \bd{k}_2 , \bd{k}_3 , \bd{k}_4 )
 \hat{P}_{ \bd{k}_1 -}  \hat{P}_{ \bd{k}_2 -}  \hat{P}_{ \bd{k}_3 -}  \hat{P}_{ \bd{k}_4 -} \Bigr)
\nonumber
 \\
 & & 
\hspace{-1mm} 
+ \frac{1}{4 !} \Bigl(
 \Gamma^{ XXXX}_{----} ( \bd{k}_1 , \bd{k}_2 , \bd{k}_3 , \bd{k}_4 )
 \hat{X}_{ \bd{k}_1 -}  \hat{X}_{ \bd{k}_2 -}  \hat{X}_{ \bd{k}_3 -}  \hat{X}_{ \bd{k}_4 -} 
%\nonumber
% \\
% & & \hspace{39mm} 
+    \Gamma^{  XXXX}_{++++} ( \bd{k}_1 , \bd{k}_2 , \bd{k}_3 , \bd{k}_4 )
 \hat{X}_{ \bd{k}_1 +}  \hat{X}_{ \bd{k}_2 +}  \hat{X}_{ \bd{k}_3 +}  \hat{X}_{ \bd{k}_4 +} \Bigr)
 \nonumber
 \\
 & &
%\hspace{39mm}     +
 \hspace{-1mm} 
+  \frac{1}{(2 !)^2} \Bigl( 
 \Gamma^{ PPPP}_{++--} ( \bd{k}_1 , \bd{k}_2 ; \bd{k}_3 , \bd{k}_4 )
 \hat{P}_{ \bd{k}_1 +}  \hat{P}_{ \bd{k}_2 +}  \hat{P}_{ \bd{k}_3 -}  \hat{P}_{ \bd{k}_4 -}
% \nonumber
% \\
% & & \hspace{39mm}    
+   
\Gamma^{ XXXX}_{--++} ( \bd{k}_1 , \bd{k}_2 ; \bd{k}_3 , \bd{k}_4 )
 \hat{X}_{ \bd{k}_1 -}  \hat{X}_{ \bd{k}_2 -}  \hat{X}_{ \bd{k}_3 +}  \hat{X}_{ \bd{k}_4 +} \Bigr)
 \nonumber
 \\
& & 
%\hspace{39mm}
 \hspace{-1mm}    
+ \frac{1}{(2 !)^2} \Bigl(
 \Gamma^{ PPXX}_{++--} ( \bd{k}_1 , \bd{k}_2 ; \bd{k}_3 , \bd{k}_4 )
 \hat{P}_{ \bd{k}_1 +}  \hat{P}_{ \bd{k}_2 +}  \hat{X}_{ \bd{k}_3 -}  \hat{X}_{ \bd{k}_4 -}
% \nonumber
% \\
%& &\hspace{39mm}     
+ 
\Gamma^{ PPXX}_{--++} ( \bd{k}_1 , \bd{k}_2 ; \bd{k}_3 , \bd{k}_4 )
 \hat{P}_{ \bd{k}_1 -}  \hat{P}_{ \bd{k}_2 -}  \hat{X}_{ \bd{k}_3 +}  \hat{X}_{ \bd{k}_4 +} \Bigr)
 \nonumber
 \\
& &  
%\hspace{39mm}
 \hspace{-1mm}  
+ \frac{1}{(2 !)^2} \Bigl(
 \Gamma^{ PPXX}_{++++} ( \bd{k}_1 , \bd{k}_2 ; \bd{k}_3 , \bd{k}_4 )
  \bigl\{ \hat{P}_{ \bd{k}_1 +}  \hat{P}_{ \bd{k}_2 +}   
 \hat{X}_{ \bd{k}_3 +}  \hat{X}_{ \bd{k}_4 +} \bigr\}
% \nonumber
% \\
%& & \hspace{39mm}  
+ 
 \Gamma^{ PPXX}_{----} ( \bd{k}_1 , \bd{k}_2 ; \bd{k}_3 , \bd{k}_4 )
  \bigl\{ \hat{P}_{ \bd{k}_1 -}  \hat{P}_{ \bd{k}_2 -}   
 \hat{X}_{ \bd{k}_3 -}  \hat{X}_{ \bd{k}_4 -} \bigr\} \Bigr)
 \nonumber
 \\
& &
%\hspace{39mm}    
+
 \Gamma^{ PXPX}_{++--} ( \bd{k}_1 ; \bd{k}_2 ; \bd{k}_3 ; \bd{k}_4 )
  \bigl\{ \hat{P}_{ \bd{k}_1 +}   \hat{X}_{ \bd{k}_2 +}  \bigr\}
 \bigl\{  
 \hat{P}_{ \bd{k}_3 -}   \hat{X}_{ \bd{k}_4 -} \bigr\}
\biggr] ,
 \label{eq:h4vertex}
 \end{eqnarray}
}
where the symmetrization symbol $ \{ \ldots \}$ is defined in
Eq.~(\ref{eq:symdef}) and we have used
{\small
 \begin{eqnarray}
 \bigl\{ \hat{P}_1 \hat{P}_2 \hat{X}_3 \hat{X}_4 \bigr\}  & = & 
 \frac{1}{2} \left[ \hat{P}_1 \hat{P}_2 , \hat{X}_3 \hat{X}_4 \right]_{+}
% \nonumber
% \\
% & + & 
+\frac{1}{4} \left( \delta_{ 1+3,0} \delta_{2+4,0} + \delta_{ 1+4,0} \delta_{2+3,0} \right).
 \hspace{7mm}
 \end{eqnarray}
}
For convenience we now introduce the short notation
$\gamma_{ \bd{k}_1 } \equiv \gamma_{ \bd{1}}$,
 $\gamma_{ \bd{k}_2 } \equiv \gamma_{ \bd{2}}$
(and similarly for the other labels) and symmetrize
the vertices whenever the interaction is symmetric with respect to the exchange
of the field labels. 
For the vertices involving four fields of the same type we obtain
{\small
\begin{eqnarray}
\Gamma^{ XXXX}_{----} ( \bd{k}_1 , \bd{k}_2 , \bd{k}_3 , \bd{k}_4 ) & = &
    \frac{ h_c}{16S}   \Big[ \gamma_{\bd{1}} + \gamma_{\bd{2}} + 
 \gamma_{\bd{3}} + \gamma_{\bd{4}} -
 2 \alpha ( \gamma_{ \bd{1} + \bd{2}}  +  \gamma_{ \bd{3} + \bd{4}}   )
+ 
( \bd{2} \leftrightarrow \bd{3} ) + ( \bd{2} \leftrightarrow \bd{4} )
\Bigr] ,\notag\\
 \label{eq:Gamma4_1}
\\
  \Gamma^{ XXXX}_{++++} ( \bd{k}_1 , \bd{k}_2 , \bd{k}_3 , \bd{k}_4 ) & = &
 \frac{ h_c}{16S}  \Bigl[ \alpha ( \gamma_{\bd{1}} + \gamma_{\bd{2}} + 
 \gamma_{\bd{3}} + \gamma_{\bd{4}} ) -
 2 \alpha ( \gamma_{ \bd{1} + \bd{2}}  +   \gamma_{ \bd{3} + \bd{4}} ) 
+  ( \bd{2} \leftrightarrow \bd{3} ) + ( \bd{2} \leftrightarrow \bd{4} )  \Bigr] ,\notag\\
 \label{eq:Gamma4_2}
 \\
\Gamma^{PPPP}_{++++} ( \bd{k}_1 , \bd{k}_2 ; \bd{k}_3 , \bd{k}_4 ) & = &
 \frac{ h_c}{16S}
 \Bigl[ - \gamma_{\bd{1}} - \gamma_{\bd{2}} - 
 \gamma_{\bd{3}} - \gamma_{\bd{4}} -
 2 \alpha ( \gamma_{ \bd{1} + \bd{2}}  +  \gamma_{ \bd{3} + \bd{4}} )    
+  ( \bd{2} \leftrightarrow \bd{3} ) + ( \bd{2} \leftrightarrow \bd{4} )
\Bigr] ,\notag\\
 \label{eq:Gamma4_5}
 \\
 \Gamma^{PPPP}_{----} ( \bd{k}_1 , \bd{k}_2 , \bd{k}_3 , \bd{k}_4 ) & = &
     \frac{ h_c}{16S}     \Bigl[ - \alpha ( \gamma_{\bd{1}} + \gamma_{\bd{2}} + 
 \gamma_{\bd{3}} + \gamma_{\bd{4}} ) -
 2 \alpha ( \gamma_{ \bd{1} + \bd{2}}  +  \gamma_{ \bd{3} + \bd{4}} )
+  ( \bd{2} \leftrightarrow \bd{3} ) + ( \bd{2} \leftrightarrow \bd{4} )
\Bigr] .\notag\\
 \label{eq:Gamma4_4}
 \end{eqnarray}
}
The vertices involving two pairs of  fields of the same type can be written as
{\small
\begin{eqnarray}
 \Gamma^{ XXXX}_{--++} ( \bd{k}_1 , \bd{k}_2 ; \bd{k}_3 , \bd{k}_4 ) & = &
 \frac{ h_c}{16S} 
 \Bigl[  \gamma_{\bd{1}} + \gamma_{\bd{2}}  + 
 \alpha ( \gamma_{\bd{3}} + \gamma_{\bd{4}} ) 
 - 2 \alpha ( \gamma_{ \bd{1} + \bd{2}}  +  \gamma_{ \bd{3} + \bd{4}} )
\Bigr] ,
 \label{eq:Gamma4_3}
 \\
 \Gamma^{ PPPP}_{++--} ( \bd{k}_1 , \bd{k}_2 ; \bd{k}_3 , \bd{k}_4 ) & = &
     \frac{ h_c}{16S}   \Bigl[ -  \gamma_{\bd{1}} - \gamma_{\bd{2}}  - \alpha 
 ( \gamma_{\bd{3}} + \gamma_{\bd{4}}) -
 2 \alpha ( \gamma_{ \bd{1} + \bd{2}}  + \gamma_{ \bd{3} + \bd{4}}  )
\Bigr] ,
 \label{eq:Gamma4_6}
\\
 \Gamma^{ PPXX}_{++--} ( \bd{k}_1 , \bd{k}_2 ; \bd{k}_3 , \bd{k}_4 ) 
& = & 
\frac{ h_c}{16S} 
 \Bigl[ 
   3 ( 
 - \gamma_{\bd{1}} -  \gamma_{\bd{2}} 
 +  \gamma_{ \bd{3}}  +  \gamma_{\bd{4}} )\notag\\
&&\hspace{8mm} -  2 \alpha \left(   \gamma_{\bd{1} + \bd{2} }  + \gamma_{\bd{3} + \bd{4} }
- \gamma_{\bd{1} + \bd{3} } - \gamma_{\bd{2} + \bd{4}} 
 -  \gamma_{\bd{2} + \bd{3} } - \gamma_{\bd{1} + \bd{4}}   \right)
\Bigr], 
 \label{eq:Gamma4_7}
\\
 \Gamma^{ PP XX}_{--++} ( \bd{k}_1 , \bd{k}_2 ; \bd{k}_3 , \bd{k}_4 )
 & = &
\frac{ h_c}{16S} 
 \Bigl[ 
   3 \alpha (
 - \gamma_{\bd{1}} -  \gamma_{\bd{2}} 
 +  \gamma_{ \bd{3}}  +  \gamma_{\bd{4}} )\notag\\
&& \hspace{8mm}-  2 \alpha \left(   \gamma_{\bd{1} + \bd{2} }  + \gamma_{\bd{3} + \bd{4} }
- \gamma_{\bd{1} + \bd{3} } - \gamma_{\bd{2} + \bd{4}} 
 -  \gamma_{\bd{2} + \bd{3} } - \gamma_{\bd{1} + \bd{4}}   \right)
\Bigr], \hspace{10mm} 
 \label{eq:Gamma4_8}
\\
 \Gamma^{ PPXX}_{++++} ( \bd{k}_1 , \bd{k}_2 ; \bd{k}_3 , \bd{k}_4 ) & = &
 \frac{ h_c}{16S} 
 \Bigl[    
- \gamma_{\bd{1}} - \gamma_{\bd{2}} + \alpha ( \gamma_{ \bd{3}}  +  \gamma_{\bd{4}} )   
- 2 \alpha (\gamma_{\bd{1} + \bd{2} } +
   \gamma_{\bd{3} + \bd{4}} ) \Bigr],
 \label{eq:Gamma4_9}
\\
 \Gamma^{ PPXX}_{----} ( \bd{k}_1 , \bd{k}_2 ; \bd{k}_3 , \bd{k}_4 ) & = &
     \frac{ h_c}{16S}   \Bigl[    
- \alpha ( \gamma_{\bd{1}} +  \gamma_{\bd{2}} ) 
 +  \gamma_{ \bd{3}}  +  \gamma_{\bd{4}}    
- 2 \alpha ( \gamma_{\bd{1} + \bd{2} } +  \gamma_{\bd{3} + \bd{4}} )
\Bigr] .
 \label{eq:Gamma4_10}
\end{eqnarray}
}
And finally, there is one vertex without permutation symmetry
connecting four different field types~\cite{footnoteminus},
{\small
 \begin{eqnarray}
 \Gamma^{ PXPX}_{++--} ( \bd{k}_1 ; \bd{k}_2 ; \bd{k}_3 ; \bd{k}_4 ) & = &
    \frac{ h_c}{16S} 
\Bigl[
  \gamma_{\bd{1}} + \alpha ( - \gamma_{\bd{2}}   
 +  \gamma_{ \bd{3}} ) - \gamma_{\bd{4}} 
-2 \alpha ( \gamma_{\bd{1} + \bd{4} } +  \gamma_{\bd{2} + \bd{3}} )
\Bigr] .
 \label{eq:Gamma4_11}
 \end{eqnarray}
}
\end{widetext}
Note that the above vertices are analytic functions of the external momenta and of $h^2$. On the other hand, if we express $\hat{H}_4^{\prime}$
in terms of the usual magnon creation and annihilation operators,
we obtain vertices which are singular for certain combinations
of external momenta \cite{Harris71,Kopietz90,Hasselmann07}.

\section{Numerical confirmation of Equation~(\ref{eq:cres1})  in two dimensions}

\renewcommand{\theequation}{B.\arabic{equation}}

In this appendix we briefly review the calculation  of the $1/S$-corrections
to the field dependent spin-wave dispersion in two dimensions as obtained within the conventional $1/S$-expansion by Zhitomirsky and Chernyshev in Ref.~[\onlinecite{Zhitomirsky99}].
From the numerical analysis of this expression we quantitatively confirm our result given in Eq.~(\ref{eq:cres1}) for the linear magnetic field dependence of the spin-wave velocity associated with the gapless magnon.
In our notation the expression for the on-shell renormalized magnon energy $\tilde E_{\bd{k}\sigma}$ given in Ref.~[\onlinecite{Zhitomirsky99}] can be written as
\begin{equation}
\tilde E_{\bd{k}\sigma}=E_{\bd{k}\sigma}+\textrm{Re } \Sigma^{1/S}_\sigma ( \bd{k} , E_{\bd{k}\sigma}+i0 
)  \, ,
\label{eq:rendisp}
\end{equation}
where the self-energy has the form
\begin{eqnarray}
\Sigma^{1/S}_\sigma ( \bd{k} , i \omega ) &=& \Sigma^{1/S}_{1\sigma} ( \bd{k} , i \omega )
+\Sigma^{1/S}_{2\sigma} ( \bd{k} , i \omega ) \nonumber \\
&& +
\Sigma^{1/S}_{3\sigma} ( \bd{k} )+\Sigma^{1/S}_{4\sigma} ( \bd{k} ) \, .
 \label{eq:sigmaZC}
\end{eqnarray}
The frequency dependent contributions to the self-energy are given by
%\begin{subequations}
\begin{eqnarray}
\hspace{-.5cm}
\Sigma^{1/S}_{1\sigma} ( \bd{k} , i \omega )=\frac{h_c^2\lambda^2}{16 S}\frac{2}{N}\sum_{\bd{q}\tau} \frac{  
\Phi_1^2(\bd{k}\sigma,\bd{q}\tau,\bd{k}-\bd{q}\overline{\sigma\tau})}{i \omega -E_{\bd{q}\tau}
-E_{\bd{k}-\bd{q}\overline{\sigma\tau}}} \, , \\
\hspace{-.8cm}
\Sigma^{1/S}_{2\sigma} ( \bd{k} , i \omega )=-\frac{h_c^2\lambda^2}{16S}\frac{2}{N}\sum_{\bd{q}\tau} \frac{
\Phi_2^2(\bd{k}\sigma,\bd{q}\tau,\bd{k}+\bd{q}\overline{\sigma\tau})}{i \omega +E_{\bd{q}\tau}
+E_{\bd{k}+\bd{q}\overline{\sigma\tau}}} \, ,
\end{eqnarray}
%\end{subequations}
where $\overline\sigma=-\sigma$ denotes a sign change such that $\overline{\sigma\tau}=-\sigma\tau$,  
and the functions $ \Phi_{1}$ and $ \Phi_{2}$ are defined as

\begin{widetext}
%\begin{subequations}
\begin{eqnarray}
%\hspace{-.8cm}
\Phi_1(\bd{k}_1\sigma_1,\bd{k}_2\sigma_2,\bd{k}_3\sigma_3)&=&
\sigma_1\gamma_{\bd{1}}(u_{\bd{1}\sigma_1}
+\sigma_1v_{\bd{1}\sigma_1})
% \nonumber \\
%&&\hspace{-3.0cm}
%\times
 (\sigma_3 u_{\bd{2}\sigma_2}v_{\bd{3}\sigma_3}+\sigma_2 u_{\bd{3}\sigma_3}v_{\bd{2}\sigma_2})
\nonumber \\ 
%&& \hspace{-3.5cm}
&+&
\sigma_2\gamma_{\bd{2}}(u_{\bd{2}\sigma_2}
+\sigma_2v_{\bd{2}\sigma_2})
% \nonumber \\
%&&\hspace{-3.0cm}
%&\times&
 (u_{\bd{1}\sigma_1}u_{\bd{3}\sigma_3}+\sigma_3 \sigma_1 v_{\bd{3}\sigma_3}v_{\bd{1}\sigma_1})
\nonumber \\ 
%&&\hspace{-3.5cm}
&+&
\sigma_3\gamma_{\bd{3}}(u_{\bd{3}\sigma_3}
+\sigma_3v_{\bd{3}\sigma_3})
% \nonumber \\
%&&\hspace{-3.0cm}
%\times
 (u_{\bd{2}\sigma_2} u_{\bd{1}\sigma_1}+
\sigma_1\sigma_2v_{\bd{1}\sigma_1}v_{\bd{2}\sigma_2}) \, ,
\\
%\end{eqnarray}
%and 
%\begin{eqnarray}
%\hspace{-.8cm}
\Phi_2(\bd{k}_1\sigma_1,\bd{k}_2\sigma_2,\bd{k}_3\sigma_3)
&=&\sigma_1\gamma_{\bd{1}}(u_{\bd{1}\sigma_1}+\sigma_1v_{\bd{1}\sigma_1}) 
%\nonumber \\
%&&\hspace{-3.0cm}
%\times
 (\sigma_2 u_{\bd{3}\sigma_3}v_{\bd{2}\sigma_2}+\sigma_3 u_{\bd{2}\sigma_2}v_{\bd{3}\sigma_3})
\nonumber \\ 
%&&\hspace{-3.5cm}
&+&
\sigma_2\gamma_{\bd{2}}(u_{\bd{2}\sigma_2}+\sigma_2v_{\bd{2}\sigma_2}) 
%\nonumber \\
%&&\hspace{-3.0cm}
%&\times&
 (\sigma_1 u_{\bd{3}\sigma_3}v_{\bd{1}\sigma_1}+\sigma_3 u_{\bd{1}\sigma_1}v_{\bd{3}\sigma_3})
\nonumber \\ 
%&&\hspace{-3.5cm}
&+&
\sigma_3\gamma_{\bd{3}}(u_{\bd{3}\sigma_3}+\sigma_3v_{\bd{3}\sigma_3}) 
%\nonumber \\
%&&\hspace{-3.0cm}
%\times
 (\sigma_1 u_{\bd{2}\sigma_2}v_{\bd{1}\sigma_1}+\sigma_2 u_{\bd{1}\sigma_1}v_{\bd{2}\sigma_2})
\, .
\end{eqnarray}
%\end{subequations}
%\end{widetext}
The frequency independent $1/S$-contributions to the self-energy are
%\begin{subequations}
%\begin{widetext}
\begin{eqnarray}
\hspace{-.0cm}\Sigma^{1/S}_{3\sigma} ( \bd{k} )&=& \frac{h_c}{2S} 
(u_{\bd{k}\sigma}^2+v_{\bd{k}\sigma}^2)[-\kappa_1 \alpha+
\kappa_2 n^2-\kappa_3 m^2 
%\nonumber \\
%&& \hspace{.9cm}
+  \sigma \gamma_{\bd{k}}(-\kappa_3 \alpha+\kappa_4 n^2/2
-\kappa_1 m^2)] 
\nonumber \\
%&& 
&-& \frac{h_c}{2S} \sigma u_{\bd{k\sigma}}v_{\bd{k\sigma}}[\kappa_2 m^2-\kappa_3 n^2 
%\nonumber \\ 
%&& \hspace{.9cm}
+2 \sigma\gamma_{\bd k}( \kappa_2 \alpha- \kappa_1 n^2+\kappa_4 m^2/2)] \, , \\
%\hspace{-.0cm}
\Sigma^{1/S}_{4\sigma} ( \bd{k} )&=&   \frac{h_c}{S} 
m^2( \kappa_2-\kappa_1+\kappa_3)
[(u_{\bd{k}\sigma}^2+v_{\bd{k}\sigma}^2)(1-\sigma\gamma_{\bd{k}}) 
%\nonumber \\
%&& \hspace{.9cm}
-2 \gamma_{\bd{k}}u_{\bd{k}\sigma}v_{\bd{k}\sigma}] \, ,
\end{eqnarray}
%\end{subequations}
\end{widetext}

  \begin{figure}[tb]    
   \centering
  \vspace{7mm}
      \epsfig{file=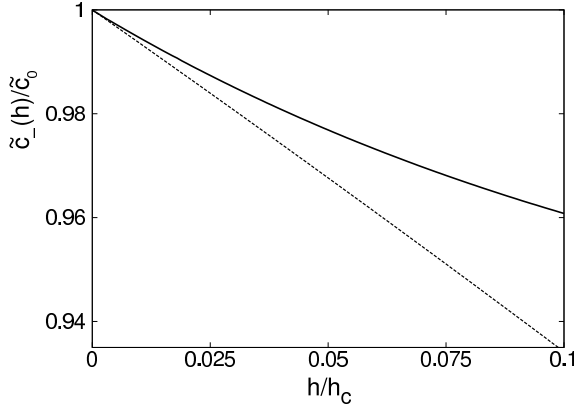,width=75mm}
  \vspace{4mm}
  \caption{%
Evolution of the spin-wave velocity of the gapless magnon as a function of the external
magnetic field for $S=1/2$. The full line shows the spin-wave velocity obtained numerically from  Eq.~(\ref{eq:rendisp})  
normalized by the zero-field value
 $\tilde c_{0} \approx 1.16 c_0$  for $S=1/2$ (see Ref.~[\onlinecite{Oguchi60}])).
The dashed line shows the prediction of Eq.~(\ref{eq:cres1}).
 Good agreement is obtained
in the limit of vanishing fields which confirms that the leading field dependence is described by 
Eq.~(\ref{eq:cres1}).
} 
    \label{fig:velocity}
  \end{figure}

with
\begin{subequations}
\begin{eqnarray}
\kappa_1 & = & \frac{2}{N}\sum_{\bd{k}\sigma} v_{\bd{k}\sigma}^2 \, , \\
\kappa_2 & = & \frac{2}{N}\sum_{\bd{k}\sigma} v_{\bd{k}\sigma} u_{\bd{k}\sigma} \gamma_{\bd{k}} \, , \\
\kappa_3 & = & \frac{2}{N}\sum_{\bd{k}\sigma} \sigma v_{\bd{k}\sigma}^2 \gamma_{\bd{k}} \, , \\
\kappa_4 & = & \frac{2}{N}\sum_{\bd{k}\sigma} \sigma v_{\bd{k}\sigma} u_{\bd{k}\sigma}  \, .
\end{eqnarray}
\end{subequations}
While the self-energy (\ref{eq:sigmaZC}) can be easily evaluated numerically, it is not
very accessible for analytical treatments and the leading small field behavior of the
spin-wave dispersion is not easily extracted from it. The equivalent
expression  Eq.~(\ref{eq:Sigmaminusres}) in the Hermitian field parametrization is more amenable
to an analytical investigation of the long wavelength physics.
To calculate the self-energy given in Eq.~(\ref{eq:sigmaZC}) we performed a two dimensional integration and used an analytical continuation to real frequencies. Performing a numerical derivative with respect to the momentum $\bd k$ at the point in the Brillouin zone where the dispersion is gapless finally yields the spin-wave velocity.
In Fig.~\ref{fig:velocity} we compare the numerically obtained spin-wave velocity
of the gapless mode at small fields with the prediction of Eq.~(\ref{eq:cres1}).
At very small fields, the numerical solution indeed confirms the behavior
given in Eq.~(\ref{eq:cres1}). For slightly larger fields, corrections beyond 
the linear dependence are also visible.

  %\vspace{-4mm}


\begin{thebibliography}{99}
  %\vspace{-4mm}
 %
\bibitem{Holstein40}    
 T.\ Holstein and H.\ Primakoff, Phys.\ Rev.\ {\bf{58}}, 1098 (1940).
%
\bibitem{Dyson56}
 F.~J.\ Dyson, Phys.\ Rev.\ {\bf{102}}, 1217 and 1230 (1956).
%
\bibitem{Maleyev57}
S. V. Maleyev, Zh. Eksp. Teor. Fiz. {\bf{30}}, 1010 (1957)
[Sov. Phys. JETP {\bf{64}}, 654 (1958)].
%
\bibitem{Oguchi60}
T. Oguchi, Phys. Rev. {\bf{117}}, 117 (1960).
%
\bibitem{Harris71}
A. B. Harris, D. Kumar, B. I. Halperin, and P. C. Hohenberg,
Phys. Rev. B {\bf{3}}, 961 (1971).
%
\bibitem{Kopietz90}
P. Kopietz, Phys. Rev. B {\bf{41}}, 9228 (1990).
%
\bibitem{Hasselmann06}
N. Hasselmann and P. Kopietz, Europhys. Lett. {\bf{74}}, 1067 (2006).
%
\bibitem{Maleyev00}
S. V. Maleyev, Phys. Rev. Lett. {\bf{85}}, 3281 (2000).
%
\bibitem{Chakravarty89}
S. Chakravarty, B. I. Halperin, and D. R. Nelson, Phys. Rev. B {\bf{39}}, 2344 (1989).
%
\bibitem{Zhitomirsky99}
M. E. Zhitomirsky and A. L. Chernyshev, Phys. Rev. Lett. {\bf{82}}, 4536 (1999).
%
\bibitem{Syromyatnikov01}
A. V. Syromyatnikov and S. V. Maleyev, Phys. Rev. B {\bf{65}}, 012401 (2001). 
%
\bibitem{Fisher89}
D. S. Fisher, Phys. Rev. B {\bf{39}}, 11783 (1989).
%
\bibitem{Sachdev99}
S. Sachdev, {\it{Quantum Phase Transitions}}, (Cambridge University Press,
Cambridge, 1999).
%
\bibitem{Schuetz03}
F. Sch\"{u}tz, M. Kollar, and P. Kopietz,
Phys. Rev. Lett. {\bf{91}}, 017205 (2003).
%
\bibitem{Spremo05}
I. Spremo, F. Sch\"{u}tz, P. Kopietz,
V. Pashchenko, B. Wolf, M. Lang, J. W. Bats, C. Hu, and M. U. Schmidt,
Phys. Rev. B {\bf{72}}, 174429 (2005).
%
\bibitem{Hasselmann07}
N. Hasselmann, F. Sch\"{u}tz, I. Spremo, and P. Kopietz,
C. R. Chimie {\bf{10}}, 60 (2007).
%
\bibitem{Anderson52}
P. W. Anderson, Phys. Rev. {\bf{86}}, 694 (1952).
%
\bibitem{Schulman81}
L. S. Schulman, {\it{Techniques and Applications of Path Integration}},
(Wiley, New York, 1981).
%
\bibitem{Negele88}
J.~W.\ Negele and H.\ Orland, {\it{Quantum Many-Particle Systems}}, (Addison-Wesley, Redwood City, 1988).
%
\bibitem{Gollisch01}
T. Gollisch and C. Wetterich, Phys. Rev. Lett. {\bf{86}}, 1 (2001);
M. Weyrauch and A. W. Schreiber,
Phys. Rev. Lett. {\bf{88}}, 078901 (2002).
\bibitem{Zhitomirsky98}
M. E. Zhitomirsky and T. Nikuni, Phys. Rev. B {\bf{57}}, 5013
(1998).
%
\bibitem{Landau80}
E. M. Lifshitz and and L. P. Pitaevskii, {\it{Statistical Physics II}},
(Pergamon, Oxford, 1980).
%
\bibitem{Kreisel07}
A. Kreisel, N. Hasselmann, and P. Kopietz, Phys. Rev. Lett. {\bf{98}}, 067203 (2007).
%
\bibitem{Castellani97}
C. Castellani, C. Di Castro, F. Pistolesi, and G. C. Strinati, Phys. Rev. Lett.
{\bf{78}}, 1612 (1997);
F. Pistolesi, C. Castellani, C. Di Castro, and G. C. Strinati, 
Phys. Rev. B {\bf{69}}, 024513 (2004).
%
% Reference of appendix:
\bibitem{Wetterich07}
C. Wetterich, arXiv:0712.1926v1.
%
\bibitem{footnoteminus}
There is a mistake in Eq.~(13e) of Ref.~[\onlinecite{Hasselmann06}]:
the term $\gamma_{\bd{1}}  - \gamma_{\bd{2}} - \gamma_{\bd{3}} + \gamma_{\bd{4}}$
should be multiplied by a factor of $2$. 
Taking into  account the different labeling of the fields in Ref.~[\onlinecite{Hasselmann06}]
as compared with the labeling in Eq.~(\ref{eq:h4vertex}) [so that we should rename $\bd{3} 
 \leftrightarrow \bd{4}$],  in the limit of vanishing magnetic field
the vertices  in Eqs.~(\ref{eq:Gamma4_1}--\ref{eq:Gamma4_11}) are then equivalent
to the vertices given in Ref.~[\onlinecite{Hasselmann06}].


  \end{thebibliography}
\end{document}